\documentclass[11pt]{article}
\usepackage{amsfonts}

\newtheorem{definition}{Definition}
\newtheorem{construction}{Construction}

\newtheorem{claim}{Claim}
\newtheorem{fact}{Fact}
\newtheorem{lemma}{Lemma}

\newtheorem{theorem}{Theorem}
\newtheorem{corollary}{Corollary}

\newcommand{\remove}[1]{}
\newcommand{\qed}{\hfill\rule{2mm}{2mm}}
\newcommand{\mod}{\mbox{\rm mod}\:}

\newcommand{\prob}{\mbox{\sf Prob}\;}

\newcommand{\bra}[1]{\langle#1\,|}
\newcommand{\ket}[1]{|\,#1\rangle}
\newcommand{\braket}[2]{\langle#1\,|\,#2\rangle}
\newcommand{\dyad}[2]{|#1\rangle\langle#2|}

\newcommand{\rank}[1]{\mbox{\sf rank}\:(#1)}
\newcommand{\fail}{\mbox{\sf FAIL}}

\newcommand{\tr}{\mbox{\rm Tr}}

\newcommand{\prefix}{\mbox{\sf LEFT}}
\newcommand{\postfix}{\mbox{\sf RIGHT}}
\newcommand{\alice}{\mbox{\sf Alice }}
\newcommand{\bob}{\mbox{\sf Bob }}

\newenvironment{proof}{\begin{trivlist}
\item[\hspace{\labelsep}{\bf\noindent Proof: }]
}{\qed\end{trivlist}}

\newcommand{\fullpaper}[1]{}

\title{Extracting Quantum Entanglement 
(General Entanglement Purification Protocols) 4}
\author{
Andris Ambainis\thanks{Supported by NSF Grant CCR-9987845 and the State of
New Jersey.}\\
School of Mathematics \\
Institute for Advanced Study \\
Princeton, NJ 08540\\
{\sf ambainis@ias.edu}
\and
Ke Yang\thanks{
A preliminary version of this paper is submitted to STOC 2002
} \\
Computer Science Department \\
Carnegie Mellon University\\
5000 Forbes Ave, Pittsburgh, PA 15213\\
{\sf yangke@cs.cmu.edu}}

\begin{document}
\maketitle


\begin{abstract}
We study the problem of general entanglement purification protocols.
Suppose Alice and Bob share a bipartite state $\rho$ which is
``reasonably close'' to perfect EPR pairs. The only information Alice
and Bob possess is a lower bound on the fidelity of $\rho$ and a
maximally entangled state.
They wish to ``purify'' $\rho$ using local operations and classical
communication and create a state that is arbitrarily close to EPR
pairs. We prove that on average, Alice and Bob cannot
increase the fidelity of the input state significantly.
We also construct protocols that may fail with a small probability,
and otherwise will output states arbitrarily close to EPR pairs with
very high probability. Our constructions are efficient, i.e., they
can be implemented by polynomial-size quantum circuits. 
\end{abstract}


\section{Introduction}


Random bits are an important 
computational resource in the randomized computation.
There has been a lot of work on extracting good random bits from
imperfect sources of randomness.
The beginning of this study goes back to at least 
von Neumann \cite{VN} who showed that a linear number of perfect random 
bits can be extracted from independent tosses of a biased coin.
More recent research has constructed
extractors \cite{NT99,Tre} which can extract almost perfect random bits 
from any source with a certain min-entropy 
without any other assumptions.
The best constructions of extractors allow to extract 
a number of random bits close to the min-entropy of the
random source with min-entropy if we can use a polylogarithmic number of 
perfect random bits \cite{TUZ}.

Quantum entanglement is an important resource in quantum computation,
similar to random bits in probabilistic computation.
It comes in the form of Einstein-Podolsky-Rosen pairs. 
An EPR pair is the state of two quantum bits 
$\frac{1}{\sqrt{2}}(\ket{00}+\ket{11})$ shared
by two parties, with one party (Alice) holding one quantum bit
and the other party (Bob) holding the second bit.
This is the quantum counterpart of a random bit shared by
two parties. 

Einstein-Podolsky-Rosen\cite{EPR35} (EPR) pairs are among the most
interesting objects of study 
in quantum mechanics and quantum information theory. 
They behave very differently from classical random bits 
shared by two parties.
The phenomenon of having \emph{entangled} 
states separated by space is one of the
quintessential features in  quantum mechanics and it has no analogue
in classical physics.  

Besides being conceptually interesting in quantum mechanics, EPR pairs
are also very useful in quantum information theory. Using an EPR
pair, Alice and Bob can perform quantum teleportation. Using
only local operations and classical communication (LOCC),
Alice can ``transport'' a qubit to Bob, who could be miles away from
Alice~\cite{BBC+93}. So EPR pairs, along with a classical
communication channel, effectively constitute a quantum
channel. Conversely, ``superdense coding'' is possible with EPR pairs:
if Alice and Bob share an EPR pair, then Alice can transport 2
classical bits to Bob by just sending one qubit~\cite{BW92}. 

For the teleportation and dense coding to work perfectly, perfect
EPR pairs are needed. Nevertheless, individual qubits are prone to
errors, 
which may
end up creating imperfect EPR pairs. These imperfect EPR pairs behave
like a noisy channel --- qubits teleported with these EPR pairs can
become distorted.

This creates the need for generating perfect (or almost perfect)
EPR pairs from imperfect ones.
This is known as  ``entanglement purification''.
Bennett et. al.~\cite{BBP+96} gave a protocol for 
the case that  Alice and Bob share identical copies of the
pure state $\ket{\phi} = (\cos\theta\ket{01} +\sin\theta\ket{10})$.
This was extended to the case when Alice and Bob share
identical copies of a mixed state \cite{BBP+96b, BDS+96,HHH96}.
Vidal~\cite{V99}, and subsequently, 
Jonathan and Plenio~\cite{JP99}, Hardy~\cite{H99},
and Vidal, Jonathan, and Nielsen~\cite{VJN00} 
considered extracting entanglement from a single copy of an arbitrary
pure state, assuming that we know a complete description of the state.


All of this work uses relatively simple 
models for imperfect EPR pairs.
The model where Alice and Bob share identical copies of the same state
corresponds to generating perfect random bits from the sequence
of i.i.d. biased coin flips.
Extracting entanglement from a single copy of a known state corresponds
to constructing uniform/almost uniform random bits 
from a biased distribution if we know a complete description
of the distribution. Both of those are very easy tasks classically
but dealing with quantum states makes them much harder.

\remove{
 were relatively simple.
In one model, Alice and Bob start with $n$ perfect
EPR pairs, and then a ``distortion'' quantum operator $\cal D$ is
applied to each pair independently, resulting in
a state $\rho$. Notice that $\cal D$ isn't necessarily a unitary
operator and the resultant state $\rho$ could be a mixed state. We
call this model the ``Identical Independent Distortion Operator''
model. The results about extracting entanglement in this model
correspond to the classical Von Neuman's result about extracting random 
bits \cite{VN} mentioned at the beginning of this paper.

A slightly different model is sometimes used, where up to $k$
of the $n$ EPR pairs are corrupted.  
The corrupted pairs can be in any state, while the rest remain
perfect. This model is often used in Quantum Error Correction
Codes, and we call it the ``Bounded Error'' model. 
These 2 models are not very different:  for example, if the operator
$\cal D$ is the depolarization channel with probability $p$, then for
some properly chosen constant $c$, at most $c\cdot np$ pairs are
corrupted with extremely high probability. By letting $k=c\cdot np$,
the two models are essentially the same. A third model is
sometimes used, where Alice and Bob share a pure state $\ket{\phi}$,
and they have complete knowledge about $\ket{\phi}$. We call this
model the ``Complete Knowledge'' model. }

Can we extract entanglement if we do not have such a detailed
knowledge of the quantum state (like extractors in the classical
setting which work for any probability distribution satisfying certain
constraints)? 
This is the question that we consider in this paper.

\subsection{Our error model}

We no longer
assume that there is a single ``distortion'' operator that acts
independently on each qubit pair, neither do we assume that Alice and
Bob have complete information about the distortion.
The only assumption that we have is that
the distortion is not very large. 
More precisely, we assume that Alice and Bob share a state $\rho$
with fidelity at least 
$(1-\epsilon)$\footnote{Most of the
  time in this paper, we are interested in the fidelity of a state
  $\rho$ and a (pre-defined) maximally entangled state (e.g., an EPR
  pair). In this case, we 
  simply use the ``fidelity of state $\rho$'' to denote the fidelity
  of $\rho$ and the pre-defined maximally entangled state of
  appropriate dimension.}.
We call this model of
imperfect EPR pairs the ``General Error'' model. 
We call the protocols for this model 
General Entanglement Purification Protocols (GEPPs)

\remove{
Aside from theoretical interest, purification of arbitrary imperfect
EPR pairs is useful in practice. Normally EPR pairs are
manufactured and transported in the 
same fashion and thus subject to the same distortion. Since the EPR
pairs can be generated independently, their distortions should be
independent.  Therefore it seems reasonable to adopt the Identical
Independent Distortion Operator model. However, it is not always true
that these distortions are really identical and independent (in many
cases, a device that starts to malfunction is more likely to continue 
malfunctioning than a device that doesn't malfunction). Furthermore,
qubits are prone to  
decoherence, and there could well be interactions between qubits when
they are stored. If such decoherence and interaction happen with
non-negligible probability, then the identical and independent
distortion assumption is no longer valid. A protocol that works with
the more adversarial General Error model is more robust than the ones
that work  with the Identical Independence Distortion Operator model.
It is a fundamental property
of quantum mechanics that qubits cannot be cloned, and that
measurements destroy general quantum states. Thus it is rather
unrealistic to assume that Alice and Bob have complete knowledge about
the state they share. }

In the General Error model, the techniques used in previous literature
don't appear to work. Some of the techniques  rely on the Law of Large
Number heavily. For example, both the ``Schmidt projection''
method~\cite{BBP+96} and the ``hashing'' method~\cite{BDS+96} try to 
reduce the state to a ``typical sequence'', and then do purification
over the typical sequences. In the General Error model, it is not clear
what a ``typical sequence'' would be. Some techniques, like the
``Procrustean'' method~\cite{BBP+96}, are designed to
work with individual states that Alice and Bob have complete knowledge
of. Apparently they don't work in the General Error model, where Alice
and Bob only have very limited information about the state they share.
In fact it is not obvious if Alice and Bob can do anything at
all to extract EPR pairs in this model.

\subsection{Our Contribution}

Some features about GEPPs are:
\begin{enumerate}
\item {\bf Arbitrary Maximally Entangled States}

  Instead of working only with EPR pairs, a GEPP works with
  arbitrary maximally entangled states. In every symmetric bipartite
  system of dimension $T\times T$, there exist a maximally entangled
  state
  $\Psi_T={1\over\sqrt{T}}\sum_{i=0}^{T-1}\ket{i}^A\ket{i}^B$. In the
  case that 
  $T$ is a power of 2, $\Psi_T$ is the state of $\log_2T$ EPR pairs.
  Thus, extracting EPR pairs is a particular case of our setting.
  We note that, by a result of Nielsen \cite{N99}, 
  Alice and Bob can transform $\Psi_N$ into $\Psi_M$ for any $M<N$.
  Therefore, if we want the final state to be a set of EPR pairs,
  we can just add a step at the end of protocol that maps $\Psi_N$ to 
  $\Psi_{2^{\lfloor \log N\rfloor}}$ and obtain a state of
  $\lfloor \log_2 N\rfloor$ EPR pairs.



\item {\bf Auxiliary Input}

  Besides an input state $\rho$, a GEPP also has a maximally
  entangled state $\Psi_K$ as auxiliary input. 
  This assumption is similar to having extra perfect random 
  bits in randomness extractors.
\remove{
  where Alice and Bob wish to 
  performance teleportation and a fidelity of $0.99$ is
  desired. Suppose Alice and Bob share $n$ near-EPR pairs of fidelity
  $0.9$ and $k$ perfect EPR pairs. Alice and Bob
  cannot directly use the $n$ near-EPR pairs for teleportation, since
  their fidelity isn't high enough. However, if there exists a GEPP
  with proper parameters, then Alice and Bob can use the $n$ near-EPR
  pairs and $k$ perfect EPR pairs to perform the GEPP and
  obtain $m$ qubit pairs of fidelity greater than $0.99$. Then Alice
  Bob can use these pairs for teleportation. So  long as $m>k$, 
  Alice and Bob will have a motivation to perform the GEPP since they
  can increase the number of ``usable'' near-EPR pairs.
}
\item {\bf Possibility to Fail}

  We allow a GEPP to fail with a reasonably low probability. As we
  will prove later, a GEPP that never fails won't be able to increase
  the fidelity of the input state significantly, even if it has an
  extra input $\Psi_K$. However, if we allow a GEPP to fail, then in
  the case that it doesn't fail, it
  will be able to output states of very high fidelity with very high
  probability. In general, a GEPP will either output a special symbol
  $\fail$ or output a state, which has (hopefully) very high fidelity.

\end{enumerate}

We consider 3 types of GEPPs.
Roughly 
speaking, we say a GEPP is \emph{absolutely successful}, if it never
fails, and always outputs a state of very high fidelity. We say a GEPP
is \emph{deterministically conditional successful}, if the probability
it fails is small, and when it doesn't fail, it will output a
state of very high fidelity with certainty. We say a GEPP is
\emph{probabilistically  conditionally successful}, if the probability
it fails is small, and in the case it doesn't fail, it outputs a state
of very high fidelity with high probability. Each definition is a
generalization of the previous one: an absolutely successful GEPP is
deterministically conditionally successful, and a deterministically
successful GEPP is probabilistically conditionally successful.
We prove the following results: 
\begin{enumerate}
\item There don't exist absolutely successful GEPPs with
  ``interesting'' parameters. To be more precise, we prove the
  following result: Suppose Alice and 
  Bob share a state of  fidelity $1-\epsilon$, and
  they have an auxiliary input $\Psi_K$. They then perform LOCC to
  create a state $\sigma$ in a subspace of dimension $M\times M$. 
  Then the maximal fidelity Alice and Bob can guarantee about the
  state $\sigma$ is at most 
  $1-{N\over N-1}(1-{K\over  M})\epsilon$. If 
  $K$ is significantly smaller than $M$ 
  improvement of fidelity 
  is very small. In other words, Alice and Bob cannot arbitrarily
  increase the \emph{average} fidelity of the input state.
\item There exist deterministically conditionally successful GEPPs for
  states in the diagonal subspace. A diagonal subspace is spanned by
  states of the form $\sum_{i}\alpha_i\ket{i}^A\ket{i}^B$. As we will
  show later, a state in the diagonal subspace is ``easy'' to work with
  and there exists an efficient protocol that is deterministically
  conditionally successful. For an input state of fidelity
  $1-\epsilon$, the protocol will fail with probability at most
  $\epsilon$, and when it doesn't fail, it always outputs a state
  of fidelity $1-{\epsilon\over L}$, where $L$ is a parameter that can
  be  made very large. We call our protocol the ``Simple Scrambling
  protocol''. The Simple Scrambling protocol is optimal in the sense
  that the average fidelity of its output matches the upper bound
  asymptotically.
\item There exist probabilistically conditionally successful GEPPs for
  arbitrary states. We present a protocol, namely the ``Hash and
  Compare protocol''. The Hash and Compare protocol converts an
  arbitrary state $\ket{\phi}$ of fidelity $1-\epsilon$ to another state
  $\ket{\phi}'$ of fidelity at least $1-\epsilon$, such that
  $\ket{\phi'}$ is ``almost'' in the diagonal subspace. Then the
  Simple Scrambling protocol can be used on state $\ket{\phi'}$ to
  create a state with very high fidelity.
\end{enumerate}

Both the Simple Scrambling protocol and the Hash and Compare protocol
are efficient, i.e., they can be implemented by polynomial-size
quantum circuits. We present the precise definitions and results in
the next section.


\section{Notations and Definitions}

\subsection{General Notations}

All logarithms are base-2. 
We use $[N]$ to
denote the set $\{0, 1, ..., N-1\}$. We identify an integer with its
binary representation, and view its binary representation as a bit
vector. The XOR of two integers $x$ and $y$, denoted by $x\oplus y$,
is the XOR of the two bit vectors $x$ and $y$ represent.
The \emph{inner product} of $x$ and $y$, denoted by $x\bullet y$, is
defined as the inner product in $GF_2$ of the 
two bit vectors $x$ and $y$ represent.

We study quantum systems
of finite dimension. We identify a pure state (written in the
``braket'' 
notation as $\ket{\phi}$) with a (column) vector of unit length. We
identify a mixed state with the density matrix of this state. 
For a quantum system whose states lie in the Hilbert space ${\cal H}$
of dimension $N$, we always assume that it has a canonical
computational basis and we denote it by $\{\ket{0}, \ket{1}, ...,
\ket{N-1}\}$. Furthermore, we often denote $\ket{0}\in{\cal H}$ by
$\ket{Z_N}$ to specify the dimension of this state.

We are mostly interested in \emph{symmetric, bipartite
quantum systems}, namely, systems shared between Alice and Bob, whose
states lie in a Hilbert space
${\cal H} = {\cal H}^A \otimes{\cal H}^B$ and 
${\cal  H}^A \equiv {\cal H}^B$.  
Alice can access ${\cal H}^A$ and Bob can access ${\cal H}^B$.
We always superscript subspaces and states to distinguish states
accessible by Alice and Bob. For example, 
a general bipartite state $\ket{\varphi}$ can written in the following
way: 
$$\ket{\varphi} = \sum_{i,j}\alpha_{ij}\ket{i}^A\ket{j}^B$$
where $\ket{i}^A$ denotes the state of Alice and
$\ket{j}^B$ denotes the state of of Bob. We sometimes subscript a
space by its dimension. For example, ${\cal H}_N$ means a space of
dimension $N$.

\remove{
Bell states refer to the following 4 states:
\begin{eqnarray*}
\Phi^+ & = & {1\over\sqrt{2}}\left(
\ket{0}^A\ket{0}^B + \ket{1}^A\ket{1}^B
\right) \\
\Phi^- & = & {1\over\sqrt{2}}\left(
\ket{0}^A\ket{0}^B - \ket{1}^A\ket{1}^B
\right) \\
\Psi^+ & = & {1\over\sqrt{2}}\left(
\ket{0}^A\ket{1}^B + \ket{1}^A\ket{0}^B
\right) \\
\Psi^- & = & {1\over\sqrt{2}}\left(
\ket{0}^A\ket{1}^B - \ket{1}^A\ket{1}^B
\right) \\
\end{eqnarray*}
These 4 states form a basis of the 2-qubit systems, and all these 4
states are maximally entangled.
}

A quantum state is unentangled if it is of the form 
$\ket{\psi}^A\otimes\ket{\psi'}^B$.
Any other pure state in ${\cal H}^A\otimes {\cal H}^B$ 
is {\em entangled}. 
For a pure state $\ket{\varphi}$ in a bipartite system, we define its
\emph{entanglement} to be the von Neumann entropy of the reduced
sub-system of Bob when we trace out Alice:
\begin{equation}
E(\ket{\varphi}) = S(\tr_A(\dyad{\varphi}{\varphi}))
\end{equation}
where $S(\rho) = -\tr(\rho\log\rho)$ is the von Neumann entropy. 
We have $S=0$ if and only the state is unentangled.  
For mixed states, a mixed state $\rho$ is unentangled if
and only if it is equivalent to a state that is a mixture of 
pure states $\ket{\varphi_i}$ with probabilities $p_i$.
Any other mixed state is entangled. However, 
there is no universally agreed definition for the amount of entanglement
in a mixed state. 

If we denote the dimension of 
${\cal H}^A$ by $N$, then the maximum amount of entanglement in this
system is $\log N$. We define the state $\Psi_N$ to be
\begin{equation}
\Psi_N = {1\over\sqrt{N}}\sum_{i=0}^{N-1}\ket{i}^A\ket{i}^B
\end{equation}
It is a maximally entangled state in 
${\cal H}^A\otimes {\cal H}^B$. Notice it is a state in a space of
dimension $N^2$. 
In particular, if $N$ is a power of 2: $N = 2^n$, then the state
$\Psi_N$ is the state of $n$ EPR pairs. We call this special kind of
states \emph{EPR states}.

\subsection{Diagonal Subspaces}

For a symmetric, bipartite system 
${\cal H} = {\cal H}_N^A\otimes {\cal H}_N^B$, we denote by 
${\cal H}^{\cal D}$ the 
$N$-dimensional subspace spanned by 
$$\left\{\;\sum_{i=0}^{N-1}\alpha_i\cdot\ket{i}^A\ket{i}^B \;\right\}$$
and we call it the \emph{diagonal subspace} of 
${\cal H}_N^A\otimes {\cal H}_N^B$. The reason for the name is: for a
general state  
$$\ket{\varphi} = \sum_{i,j}\alpha_{ij}\ket{i}^A\ket{j}^B$$
we can write its coefficients (totally $N^2$ of them) in a matrix
form, where the $(i,j)$-th entry is $\alpha_{i,j}$, then the elements
in ${\cal H}^{\cal D}$ correspond to the diagonal matrices. Notice
that this definition is also consistent with the 
``Bell-diagonal''~\cite{BDS+96} states for $N=2$. 
A mixed state $\rho$ is in the diagonal subspace, if there exists a
decomposition of $\rho$:
$$\rho = \sum_{i}p_i\cdot \dyad{\phi_i}{\phi_i}$$
such that  all pure states $\ket{\phi_i}$ are in the diagonal subspace.

\subsection{Fidelity}
\label{section-fidelity}

For two (mixed) states $\rho$ and $\sigma$ in the same quantum system,
their \emph{fidelity} is defined as
\begin{equation}
\label{eqn:def-fidelity-orig}
F(\rho, \sigma)=\tr(\rho^{1/2}\sigma\rho^{1/2}).
\end{equation}
This definition simplifies if one 
$\sigma = \dyad{\varphi}{\varphi}$ is a pure state. Then the fidelity of
$\rho$ and $\sigma$ is
\begin{equation}
\label{eqn:def-fidelity-pure}
F(\rho, \dyad{\varphi}{\varphi})=
\bra{\varphi}\rho\ket{\varphi}
\end{equation}
In the special case that $\ket{\varphi} = \Psi_N$ is the
maximally entangled state, we call the fidelity of $\rho$ and
$\ket{\varphi}$ the \emph{fidelity
  of state $\rho$}, and the definition simplifies to:
\begin{equation}
\label{eqn:def-fidelity-new}
F(\rho) = \bra{\Psi_N}\rho\ket{\Psi_N}
\end{equation} 
When the state $\rho=\dyad{\phi}{\phi}$ is also a pure state, we have
$$F(\rho) = F(\dyad{\phi}{\phi}) = |\braket{\phi}{\Psi_N}|^2$$
and the fidelity is just the square of the inner product with $\Psi_N$.

One property for the fidelity is: it is linear with respect to
ensembles. 
\begin{claim}
\label{claim:linear-fidelity}
Let $\rho$ be the density matrix for a mixed state that is an ensemble
$\{p_i,\;\ket{\phi_i}\}$. The fidelity of $\rho$ is the weighted
averages of the qualities of the pure states:
$$F(\rho) = \sum_{i}p_i\cdot F(\dyad{\phi_i}{\phi_i})$$
\qed
\end{claim}
This linearity is particularly convenient in some of the proofs in
this paper.

\section{General Entanglement Purification Protocols}

\subsection{The general setting}

Alice and Bob are given some entangled state 
in ${\cal H}_N^A\otimes{\cal H}_N^B$.
They are also given an auxiliary input 
$\Psi_K$ in ${\cal H}_K^A\otimes{\cal H}_K^B$.
Alice can perform unitary transformations on her part of the state
(${\cal H}_N^A \otimes {\cal H}_K^A$) and
Bob can perform unitary transformations on his part
(${\cal H}_N^B \otimes {\cal H}_K^B$).
Since those transformations only affect one part of the
state, they are called {\em local operations}.
Alice and Bob are also allowed to communicate classical bits but not
quantum bits.
This model is called {
\em LOCC (local operations and classical communication)}
\cite{BBP+96,N99}.

If the starting state is unentangled, applying LOCC 
operations keeps the state unentangled \cite{BBP+96}.
Thus, LOCC operations cannot create entanglement but
they can be used to extract the entanglement that
already exists in the state.

We use the letter ${\cal P}$ to denote protocols
for extracting entanglement by LOCC operations.
At the end of a protocol ${\cal P}$, \alice and \bob have two options: 
\begin{enumerate}
\item They can abort and claim failure by outputting a special
  symbol $\fail$. We denote this by 
${\cal P}(\rho) = \fail$.
\item They can
output a (possibly mixed) state $\sigma$ in 
${\cal H}^A_{M}\otimes{\cal  H}^B_{M}$.  We denote this by 
${\cal P}(\rho) = \sigma$.
\end{enumerate}

We now define the error model.
We first give an unsuccessful definition to illuminate some
of difficulties that we face and to explain the reasons
behind our final definition.

\subsection{Extracting entanglement from an arbitrary state}

Ideally, we would like to have a protocol that takes any
entangled state in ${\cal H}^A_N\otimes{\cal H}^B_N$
with at least a certain amount of entanglement and
extracts a state close to $\Psi_M$ for some $M<N$.
This would correspond the definition of extractors where extractor
can transform any probability distribution with min-entropy 
at least $m$ into a probability distribution that is
close to uniform.

Unfortunately, this is not possible, even  
if we restrict ourselves to starting states with the maximum possible 
entanglement. Unlike in the classical world where there is just
one probability distribution over $N$ elements with entropy $\log N$
(the uniform distribution), there are infinitely many quantum
states with entanglement $\log N$. Namely, any quantum state
of the form 
\begin{equation}
\label{eq-andris-maxent} 
\ket{\phi} = \sum_{i=0}^{N-1} \alpha_i \ket{i}\ket{i} 
\end{equation}
with $|\alpha_i|^2 = 1/N$ for all $i\in\{0, \ldots, N-1\}$
has entanglement $\log N$.
In particular, this includes 
\[ \ket{\phi_j} = 
\sum_{i=0}^{N-1} \frac{1}{\sqrt{N}} e^{2 ij\pi/N} \ket{i}\ket{i} \]
for $j\in\{1, \ldots, N\}$.
Assume that we have a protocol that 
extracts $\Psi_M$ from any $\ket{\phi_j}$.
This means that, given $\ket{\phi_j}$, the protocol ends
with the final state of the from $\Psi_M \otimes \ket{\phi'_j}$.
We consider running this protocol on the mixed state $\rho$
that is $\ket{\phi_0}$ with probability 1/N, $\ket{\phi_2}$
with probability 1/N, ..., $\ket{\phi_{N-1}}$ with probability 1/N.
Then, the final state is of the form $\Psi_M \otimes \rho'$
where $\rho'$ is some mixed state.

The problem is that $\rho$ is equivalent to the mixed state
that is $\ket{0}\ket{0}$ with probability $1/N$,
$\ket{1}\ket{1}$ with probability $1/N$, ...,
$\ket{N-1}\ket{N-1}$ with probability $1/N$.
(This equivalence can be verified by writing out the density
matrices of both states.)
Neither of states $\ket{i}\ket{i}$ is entangled,
so the mixed state obtained by combining them is not
entangled as well. Yet, since this mixed state is
equivalent to $\rho$, it gets transformed into $\Psi_M\otimes \rho'$
which is entangled because $\Psi_M$ is entangled.

We have constructed a protocol that transforms an unentangled starting
state into entangled end state without quantum communication.
Since this is impossible \cite{BBP+96},
our assumption is wrong and there is no
protocol that extracts any $\Psi_M$ from an arbitrary $\ket{\phi_j}$.

The argument described above is still valid if we relax the requirement
to extracting a state close to $\Psi_M$ and
if we allow to use a perfect auxiliary state $\Psi_K$. 
In the second case, we can get the perfect $\Psi_K$ back
but cannot get an entangled state of higher dimension.

\subsection{Extracting from a state close to $\Psi_M$}

The reason for the problem in the previous section is that 
there are multiple maximally entangled states and combining
them into a mixed state can cancel the entanglement and 
create a state with no entanglement.
The consequence is that if we want to be able to extract entanglement
we have to restrict ourselves to states that are close to one particular
highly entangled state (rather than some highly entangled state).
Therefore, we assume that the starting
state is close to $\Psi_M$\footnote{The protocols can be modified 
to use any other fixed state of the form (\ref{eq-andris-maxent}) instead of $\Psi_M$}. 

A common way to measure the closeness to $\Psi_M$ is the fidelity (section
\ref{section-fidelity}). This gives the following definitions.

\begin{definition}[Absolutely Successful GEPP]
\label{def:abs-gepp}
A General Entanglement Purification Protocol ${\cal P}$ is
\emph{absolutely successful with parameter}  
$\langle N, K, M, \epsilon, \delta\rangle$, if for all states 
$\rho$ such that  $F(\rho)\ge 1-\epsilon$, 
$$\prob[{\cal P}(\rho) = \fail] = 0$$ and
$$\prob[F({\cal P}(\rho)) \ge 1-\delta] = 1$$
\end{definition}

\begin{definition}[Deterministically Successful GEPP]
\label{def:det-cond-gepp}
A General Entanglement Purification Protocol ${\cal P}$ is
\emph{deterministically conditionally successful with
  parameter} $\langle N, K, M, \epsilon, \delta, p\rangle$,
if for all input states $\rho$ such that $F(\rho) = 1-\epsilon$, 
$$\prob[{\cal P}(\rho) = \fail] \le p$$
and
$$\prob[F({\cal P}(\rho)) \ge 1-\delta\;|\;{\cal P}(\rho) \ne \fail] =
1$$
\end{definition}

\begin{definition}[Probabilistically Successful GEPP]
\label{def:prob-cond-gepp}
A General Entanglement Purification Protocol ${\cal P}$ is
\emph{probabilistically conditionally successful with
  parameter} $\langle N, K, M, \epsilon, \delta, p, q \rangle$,
if for all input states $\rho$ such that $F(\rho) = 1-\epsilon$, 
$$\prob[{\cal P}(\rho) = \fail] \le p$$
and
$$\prob[F({\cal P}(\rho)) \ge 1-\delta\;|\;{\cal P}(\rho) \ne \fail]
\ge 1-q$$
\end{definition}

\begin{definition}[Efficient GEPP]
\label{def:eff-gepp}
A General Entanglement Purification Protocol ${\cal P}$ is
\emph{efficient}, if there exists a constant $c$ such that ${\cal P}$
can be implemented by quantum circuits of size 
$O((\log N + \log K)^c)$.
\end{definition}

\section{Results}


\subsection{Impossibility result for absolutely successful protocols}

\begin{theorem}
\label{thm:succ-bound}
\begin{enumerate}
\item[{\bf (a)}]
For all absolutely successful General Entanglement Purification
Protocols with parameter $\langle N, K, M, \epsilon, \delta \rangle$,
we have the following inequality:
$$\delta \ge {{M-K}\over M}{N\over {N-1}} \epsilon.$$
\item[{\bf (b)}]
The bound in (a) is tight. For any integers $N, M, K$ such that
$NK/M$ and $M/K$ are both
integers, there exists an absolutely successful GEPP with parameter
\begin{math}
\langle N, K, M,\epsilon, {{M-K}\over M}{N\over {N-1}} \epsilon
\rangle
\end{math}.
\end{enumerate}
\end{theorem}

This shows that absolutely succesful protocols are quite weak.
If we just want to extract the auxiliarly state $\Psi_K$ and 
$c$ more EPR pairs, then $M=2^c K$ and we can achieve
the fidelity of at most $1-{{2^c-1}\over {2^c}}{N\over {N-1}}\epsilon<
1-(1-{1\over 2^c}) \epsilon$ which is
less than ${1\over 2^c}\epsilon$ better than $1-\epsilon$ that we had
at the beginning. If we want to get $\Psi_K$ plus
a linear number of EPR pairs, the improvement in fidelity is
an exponentially small fraction of $\epsilon$.

We prove theorem \ref{thm:succ-bound} in appendix \ref{appendix-1}.

\subsection{Constructions of conditionally successful protocols}

On the other hand, there are good conditionally successful protocols.

\begin{theorem}
\label{thm:succ-det}
For all integers $n, t, d$ such that $n>t$ and any real 
$\epsilon<1/2$, there exist 
efficient deterministic conditionally successful general entanglement
purification protocols of following parameters
\begin{itemize}
\item
\begin{math}
\langle 2^n, (2^n-1), 2^{n-t}(2^n-1), \epsilon, {\epsilon\over{2^t}},
\epsilon
\rangle
\end{math}
\item
\begin{math}
\langle 2^{2n}, 2^n+1, 2^{n}(2^n+1), \epsilon,
{\epsilon\over{2^{n}}}, \epsilon 
\rangle
\end{math}
\item
\begin{math}
\langle 2^{dn}, {{2^{dn}-1}\over{2^n-1}},
2^{(d-1)n}({{2^{dn}-1}\over{2^n-1}}), 
\epsilon, 
{\epsilon\over{2^{n}}}, \epsilon
\rangle
\end{math}
\end{itemize}
for mixed states in the diagonal subspace.
\end{theorem}

The first protocol achieves the smallest loss of EPR pairs,
increasing the fidelity from $1-\epsilon$ to $1-{\epsilon\over 2^t}$
at the cost of losing just $t$ EPR pairs. Is starts with $n$ 
imperfect EPR pairs and an auxiliary state of dimension $2^n-1$ and
outputs a state of dimension $2^{n-t} (2^n-1)$.
The disadvantage is that we have to use an auxiliary state of
almost the same dimension ($2^n-1$) as the state that we try to
purify ($2^n$).
The second and the third construction use smaller auxiliary states
but lose more EPR pairs.
All 3 results are achieved by 
Simple Scrambling Protocol (appendix \ref{appendix-2})
using 3 different constructions of scrambling permutations (appendix
\ref{sec:construct-scramble}).

The constructions fail with probability at most $\epsilon$. 
We can extend Theorem \ref{thm:succ-det} to show that
the trade-off between the probability of failure and 
increase in fidelity achieved by Theorem \ref{thm:succ-det}
is optimal. It might be possible to improve the theorem
with respect to other parameters (the dimensionality of
the extra state $\Psi_K$ and the amount of entanglement
that is lost if the prototocol does not fails).

For states not in the diagonal subspace, we can construct a probabilistically
succesful protocol with almost the same parameters.

\begin{theorem}
\label{thm:complete-scramble}
For all integers $n, t, l, d$ such that $n>t$, $n>l$ and all real
$\epsilon<1/2$, there exist 
efficient probabilistic conditionally successful general entanglement
purification protocols of following parameters
\begin{itemize}
\item
\begin{math}
\langle 2^n, (2^n-1)2^{2t}, 2^{n-t}(2^n-1), \epsilon, {\epsilon\over{2^{t-3}}},
2\epsilon + \sqrt{2\epsilon\over{2^t}}, {1\over{2^t}} \rangle
\end{math}
\item
\begin{math}
\langle 2^{2n}, (2^n+1)2^{2t}, 2^{n}(2^n+1), \epsilon,
{\epsilon\over{2^{t-3}}}, 2\epsilon 
+ \sqrt{2\epsilon\over{2^t}}, {1\over{2^t}} \rangle
\end{math}
\item
\begin{math}
\langle 2^{dn}, {{2^{dn}-1}\over{2^n-1}}\cdot 2^{2t}, 
2^{(d-1)n}({{2^{dn}-1}\over{2^n-1}}), 
\epsilon, {\epsilon\over{2^{t-3}}}, 2\epsilon
+ \sqrt{2\epsilon\over{2^t}}, {1\over{2^t}} \rangle
\end{math}
\end{itemize}
\end{theorem}

We note that the extra probability of failure ($\sqrt{2\epsilon\over{2^t}}$)
can be made arbitrarily small by increasing $t$.
This theorem is shown by Complete Scrambling protocol which combines the Simple Scrambling protocol
with another protocol, Hash-and-Compare (appendix \ref{appendix-3}).

\section{Conclusions and Open Problems}

We investigated the problem of entanglement purification by
Alice and Bob via LOCC. We used a very general model of the input
state, where the only information Alice and Bob have is a lower bound
on the fidelity of the input state. This contrasts with the previous models
which assumed that the ``noise'' is identical
and independent, or Alice and Bob have complete knowledge of the input
state they share. Because of the generality of the General Error
model, the techniques used in previous works don't appear viable.

We defined three types of General Entanglement Purification Protocols.
Absolutely successful GEPPs never fail, and
they always output states that have very high fidelity.
Deterministically conditionally successful GEPPs fail with small
probability, and otherwise, they output states of high fidelity with
certainty.
Probabilistically conditionally successful GEPPs fail with
small probability, and otherwise output states of high fidelity with
very high probability.

We proved a negative result that there don't exist absolutely 
successful GEPPs of interesting parameters, i.e., on average, the
ability of Alice and Bob to purify the entanglement is very limited.

We constructed efficient GEPPs that are deterministically
conditionally successful for mixed states in the diagonal subspace (Simple Scrambling
protocol) and probabilistically conditionally successful for 
an arbitrary state of sufficiently high fidelity 
(Complete Scrambling protocol).

\remove{
We proved this by demonstrating a construction of the Simple Scrambling
protocol. This protocol is
efficient if the Scrambling Permutation it uses is efficient. 

We presented a construction of the Hash and Compare protocol. This
protocol converts a state of reasonably high fidelity into a state of
the same or higher fidelity, which is ``almost completely'' in the
diagonal subspace. The Hash and Compare protocol is efficient.

Finally, we combine the Simple Scrambling protocol and the
Hash and Compare protocol to obtain the Complete Scrambling protocol,
which is probabilistically conditionally successful for any state of
reasonably high fidelity.}

In our construction of the protocols, Scrambling Permutations play a
very important role. We give 3 different constructions of efficient
Scrambling Permutations in Appendix~\ref{sec:construct-scramble}. Each
construction has its own advantage. By plugging
them 
into the
construction of Complete Scrambling protocol, we obtain different
protocols with different parameters. We notice
that the notion of Scrambling Permutations are closely related to
universal hash functions. By being more lax on the ``scrambling''
property, they can have more efficient constructions than 
universal hash functions.

There are several open problems:
\begin{enumerate}

\item{\bf Remove the auxiliary input or reduce its size.}
In our paper, both the Simple Scrambling protocol and the Hash and
Compare protocol need maximally entangled states as auxiliary input.
The Simple Scrambling protocol needs them to ``scramble'' the
coefficients of the input state, and the Hash and Compare protocol
needs them to perform teleportation. In the final construction of
the Complete Scrambling protocols, if
Alice and Bob share an input of $n_0$ qubit pairs, they need to invest
$O(n_0)$ perfect EPR pairs in order to perform the
purification. It would be very desirable to reduce the number of
auxiliary perfect EPR pairs as much as possible: the ideal case
would be removing them completely, but even reducing them to $o(n_0)$
would be interesting.

\item{\bf Optimal GEPPs}
The Simple Scrambling protocol is optimal in the sense that the
average fidelity of its output matches the upper bound of
Theorem~\ref{thm:succ-bound} asymptotically (in case Alice and Bob
fail, they can output $\ket{Z_M}^A\otimes \ket{Z_M}^B$
instead). However it is not true for Hash and Compare 
and Complete Scrambling protocols. Do
there exist optimal GEPPs that work for all states?

\item{\bf Relationship to classical randomness extraction.}
As we described in the introduction, we view the problem of
extracting entanglement as a quantum counterpart of extracting
randomness from a weak random source.
There is also a similarity between the techniques used in 
those two problems.
One of the main techniques
used in the classical randomness extraction is the universal hash
function, and we used scrambling permutations in our construction of
GEPP. 
Are there deeper
relationships between the two problems? Also, 
can some of the techniques from classical randomness extractors 
be used in the entanglement purification? Notice
that the state of art in randomness extraction is that only
logarithmic number of truly random bits need to be invested and
almost all the entropy can be extracted~\cite{NT99}, 
whereas in the case
of entanglement extraction, our constructions call for linear number
of perfect EPR pairs to be invested and considerable amount of
entanglement is wasted. Can we make the entanglement purification
protocols more efficient, or are these inefficiencies inherent?

\end{enumerate}

\section*{Acknowledgment}
The authors would like to thank Bob Griffith and Steven Rudich for
enlightening discussions. The authors thank John
Langford for proof-reading the earlier version of this paper.


\newpage
\appendix


\section{Tight bounds on Absolutely Successful GEPPs}
\label{appendix-1}

We prove Theorem~\ref{thm:succ-bound} in this section.

\subsection{A Negative Result}

We show that on
average, Alice and Bob cannot increase the fidelity of their input
state significantly, even if they have an auxiliary input $\Psi_K$.

We first study a simpler problem. 
Suppose Alice and Bob share a maximally entangled state $\Psi_K$ and
some private ancillary bits, initialized to $\ket{0}$. We describe this
shared state by 
$$\ket{\phi} = (\ket{Z_{N}}^A\otimes\ket{Z_N}^B)\otimes\Psi_K$$
The fidelity of this state is $K/M$ by a simple computation.

Alice and Bob try to convert state $\ket{\phi}$ as close to
$\Psi_{M}$ as possible by LOCC. The problem is: how close can they get?
If $M/K$ is an integer, Alice and Bob just trace out a
subsystem of their ancillary bits to bring the dimension of each their
subsystem to 
$M$, then they obtain a state 
$$\ket{\psi_0} =  (\ket{Z_{M/K}}^A\otimes\ket{Z_{M/K}}^B)\otimes\Psi_K$$
which has fidelity $K/M$ by a straightforward computation. In fact,
this is actually the best Alice and Bob can do:

\begin{lemma}
\label{lemma:negative-simple-bi}
Let $\ket{\phi} = (\ket{Z_N}^A\otimes\ket{Z_N}^B)\otimes\Psi_K$ be a
state in a bipartite system ${\cal H}_{NK}^A\otimes{\cal H}_{NK}^B$
shared between Alice and Bob. Let $\sigma$ be the state Alice and Bob
output after performing LOCC operations. Suppose that $\sigma$ is in
the subspace ${\cal H}_{M}^A\otimes{\cal H}_{M}^B$. We have
$F(\sigma) \le {K\over M}$. 
\qed
\end{lemma}

This lemma is a direct corollary of a result by Vidal, Jonathan, and
Nielsen~\cite{VJN00}. 
There is also a simple direct proof in 
Appendix~\ref{sec:proof-neg-simple}.

\begin{proof}[Proof to Theorem~\ref{thm:succ-bound}, part (a)]

We prove the theorem by demonstrating a particular mixed state $\rho$
such that $\rho$ has a fidelity $1-\epsilon$, and no LOCC can increase
its fidelity to more than $1- {{M-K}\over M}{N\over {N-1}} \epsilon$.

Let $\epsilon'=\frac{N}{N-1}\epsilon$.
We define the state $\rho$ to be
$$\rho = (1-\epsilon')\cdot\dyad{\Psi_N}{\Psi_N} +
\epsilon'\cdot\dyad{Z_N^A\otimes Z_N^B}{Z_N^A\otimes Z_N^B}$$
In fact, $\rho$ is the maximally entangled state
$\Psi_M$ with probability $(1-\epsilon')$ and the totally disentangled
state $Z_N^A\otimes Z_N^B$ with probability $\epsilon'$.  

It is easy to verify that
$F(\rho) = 1-\epsilon$, since 
$\braket{\Psi_N}{Z_N^A\otimes Z_N^B} = 1/\sqrt{N}$
and, therefore,
$$ F(\rho)=(1-\epsilon') F(\dyad{\Psi_N}{\Psi_N}) + \epsilon'
F(\dyad{Z_N^A\otimes Z_N^B}{Z_N^A\otimes Z_N^B}) = 
(1-\epsilon')+{1\over N}\epsilon' = 1 - (1-{1\over N}) \epsilon' = 
1-\epsilon . $$

For an arbitrary GEPP ${\cal P}$ that never fails, we define 
$$f_1 = F({\cal P}(\dyad{\Psi_N}{\Psi_N}))$$ and
$$f_2 = F({\cal P}(\dyad{Z_N^A\otimes Z_N^B}{Z_N^A\otimes Z_N^B}))$$
Then we have $f_1 \le 1$ and by  Lemma~\ref{lemma:negative-simple-bi},
$f_2\le K/M$.

By the linearity of fidelity of quantum operations, we know that
$$F({\cal P}(\rho)) = (1-\epsilon')f_1 + \epsilon' f_2 \le 
1-{{M-K}\over M}\epsilon'= 1 -{{M-K}\over M}{N \over {N-1}}
\epsilon.$$

We will prove the part (b) of the theorem  in the next subsection.
\end{proof}

Therefore, there don't exist absolutely successful GEPPs
with very interesting parameters --- we hope
that our protocol is able to ``boost'' the fidelity of the input state
to arbitrarily close to 1, but clearly this is impossible for
absolutely successful protocols.

\subsection{A Protocol That Matches the Bound}

We now prove the second part Theorem~\ref{thm:succ-bound}, that
the bound is tight. We do so by showing that there is a protocol that
achieves this slight increase in fidelity. 

The input to the protocol is a state $\rho$ in 
${\cal  H}_N^A\otimes{\cal H}_N^B$. 
The protocol outputs a state in ${\cal H}_M^A\otimes{\cal H}_M^B$
where $M<N$ and $M$ divides $N$.
There is no auxiliary state used, i.e., $K=1$.

\begin{construction}[Random Permutation Protocol]
\label{construct:absolute-scramble}
The input to this protocol is a state in 
${\cal H}_N^A\otimes {\cal H}_N^B$.  The steps are:

\begin{enumerate}
\item Alice generates a uniformly random permutation $\pi$ on $N$
  elements 
using classical randomness and transmits the permutation to Bob.
\item
Alice applies permutation $\pi$ on ${\cal H}_N^A$,
mapping $\ket{i}$ to $\ket{\pi(i)}$,
Bob does the same on ${\cal H}_N^B$.
\item
Alice and Bob decompose ${\cal H}_N$ as ${\cal H}_M\otimes {\cal H}_L$,
$L=N/M$ and measure the ${\cal H}_L$ part. 
\item
Alice sends the result of her measurement to Bob, Bob sends
his result to Alice.
\item
They compare the results. If the results are the same, they output
the state that they have in ${\cal H}_M^A\otimes {\cal H}_M^B$.
If the results are different, they output $\ket{Z_M} \otimes \ket{Z_M}$.
\end{enumerate}
\end{construction}

We start with the case when the state of Alice and Bob is in the
diagonal subspace. 

\begin{lemma}
\label{lem:asp-diagonal}
If the input state to the Random Permutation Protocol is 
in the diagonal subspace, the protocol is absolutely
successful with parameters
$\langle N, 1, M, \epsilon, {{M-1} \over M}{N\over {N-1}}\epsilon\rangle$.
\end{lemma}

\begin{proof}
Without loss of generality, we assume that the starting state is pure.
Let $\ket{\phi}=\sum_{i=1}^N \alpha_i \ket{i}^A\ket{i}^B$ be the
starting state. 
For a permutation $\pi$, let $U_{\pi}$ be the unitary transformation
defined 
by 
\begin{math}
U_{\pi}\left(\ket{i}^A\otimes\ket{j}^B\right)=
\ket{\pi(i)}^A\ket{\pi(j)}^B
\end{math}.
Then, if Alice and Bob use a permutation $\pi$,
the resulting state is 
$$\ket{\phi_\pi}=U_\pi \ket{\phi}  = 
\sum_{i=1}^N \alpha_i \ket{\pi(i)}^A\ket{\pi(i)}^B = 
\sum_{i=1}^N \alpha_{\pi^{-1}(i)} \ket{i}^A\ket{i}^B.
$$

There are $N!$ permutations $\pi$ on a set of $N$ elements.
Therefore, each of them gets applied with probability $1/N!$.
This means that the final state is a mixed state of $\ket{\phi_\pi}$
with probabilities $1/N!$ each.
We calculate the density matrix $\rho$ of this state.
It is equal to 
\[ \sum_{\pi} {1\over N!} \ket{\phi_\pi}\bra{\phi_\pi} =
\sum_{\pi} {1\over N!} \left( 
\begin{array}{llll}
\alpha_{\pi^{-1}(1)}\alpha^*_{\pi^{-1}(1)} & 
\alpha_{\pi^{-1}(1)}\alpha^*_{\pi^{-1}(2)} & \ldots &
\alpha_{\pi^{-1}(1)}\alpha^*_{\pi^{-1}(N)} \\
\alpha_{\pi^{-1}(2)}\alpha^*_{\pi^{-1}(1)} & 
\alpha_{\pi^{-1}(2)}\alpha^*_{\pi^{-1}(2)} & \ldots &
\alpha_{\pi^{-1}(2)}\alpha^*_{\pi^{-1}(N)} \\
\ldots & \ldots & \ldots & \ldots \\
\alpha_{\pi^{-1}(N)}\alpha^*_{\pi^{-1}(1)} & 
\alpha_{\pi^{-1}(N)}\alpha^*_{\pi^{-1}(2)} & \ldots &
\alpha_{\pi^{-1}(N)}\alpha^*_{\pi^{-1}(N)} 
\end{array}
\right) .\]
We claim that all diagonal entries $\rho_{ii}$ are equal to $1/N$ and
all off-diagonal entries $\rho_{ij}$, $i\neq j$ are equal to some value $a$
which is real. This follows from the symmetries created by summing over all
permutations.

Consider a diagonal entry $\rho_{ii}$. 
For each $j\in\{1, \ldots, N\}$, there are $(N-1)!$ permutations
that map $j$ to $i$. Therefore,
\[ \rho_{ii}= \sum_{j=1}^N (N-1)! {1\over N!} 
\alpha_{j}\alpha^*_{j} = 
{1\over N} \sum_{j=1}^N |\alpha_{j}|^2 .
\]
$\sum_{j=1}^N |\alpha_j|^2$ is the same as $\|\phi\|^2$ which is equal to 1.
Therefore, $\rho_{ii}={1\over N}$.

Next, consider an off-diagonal entry $\rho_{ij}$. 
For each $k, l$, $k\neq l$, there are $(N-2)!$ permutations
that map $k$ to $i$ and $l$ to $j$. Therefore,
\[ \rho_{ij}= \sum_{k=1}^N \sum_{l=1, l\neq k}^N (N-2)! {1\over N!} 
\alpha_{k}\alpha^*_{l} = 
\sum_{k=1}^N \sum_{l=1, l\neq k}^N {1\over N(N-1)}
\alpha_{k}\alpha^*_{l} .\] 
This immediately implies that $\rho_{ij}$ is the same for all $i\neq j$.
Also, notice that $(\alpha_k\alpha^*_l)^* = \alpha^*_k \alpha_l$.
Therefore, $\alpha_k\alpha^*_l+\alpha_l \alpha^*_k$ is real
and $\rho_{ij}$ (which is a sum of terms of this form) is real as well.
Let $a=\rho_{ij}$.
We have shown that 
\[ \rho = \left(
\begin{array}{llll}
{1\over N} & a & \ldots & a \\
a & {1\over N} & \ldots & a \\
\ldots & \ldots & \ldots & \ldots \\
a & a & \ldots & {1 \over N} 
\end{array} \right) . \]

Notice that
the density matrix $\rho$ can be also obtained from a mixed state
that is $\Psi_N$ with probability $N a$ and each of basis states 
$\ket{i}^A\ket{i}^B$ with probability $\frac{1}{N}-a$.

We now consider applying steps 3-5 to those states.
Measuring ${\cal H}_L^A\otimes {\cal H}_L^B$ 
for $\ket{\Psi_N}$ always gives the 
same results and leaves Alice and Bob with
the state $\ket{\Psi_M}$ in ${\cal H}_M^A\otimes{\cal H}^B_M$.
The fidelity of this state with $\ket{\Psi_M}$ is, of course, 1.
Measuring ${\cal H}_L^A$ and ${\cal H}_L^B$ for $\ket{i}^A \ket{i}^B$ also
gives the same results and leaves Alice and Bob with some basis
state $\ket{i'}^A \ket{i'}^B$ in the diagonal subspace of 
${\cal H}_M^A\otimes {\cal H}_M^B$.
The fidelity of this state and $\ket{\Psi_M}$ is $\frac{1}{M}$.
By Claim \ref{claim:linear-fidelity}, 
if we apply those steps to the state $\rho$,
we get that the final fidelity 
\begin{equation}
\label{eq-andris1} 
N a + N \left( {1\over N}-a\right) {1\over M} = 
{1\over M} + N a \left( 1-{1\over M} \right) .
\end{equation}

We now lower-bound $a$. 
By Claim \ref{claim:linear-fidelity}, $F(\rho)={1\over N!} \sum_{\pi}
F(\ket{\phi_\pi})$.
Since permuting the basis states $\ket{i}^A\ket{i}^B$ 
preserves the maximally entangled 
state $\Psi_N={1\over \sqrt{N}}\sum_{i=1}^N \ket{i}^A\ket{i}^B$,
the fidelity of any $\ket{\phi_{\pi}}$ is the same as
the fidelity of $\ket{\phi}$.
Therefore, $F(\rho)=F(\ket{\phi})\geq 1-\epsilon$.
By applying the definition of fidelity,
\[ F(\rho) = \left( \begin{array}{l} {1\over\sqrt{N}} \\ 
 {1\over\sqrt{N}} \\ \ldots \\{1\over\sqrt{N}} \end{array} \right)
 \left(
\begin{array}{llll}
{1\over N} & a & \ldots & a \\
a & {1\over N} & \ldots & a \\
\ldots & \ldots & \ldots & \ldots \\
a & a & \ldots & {1 \over N} 
\end{array} \right)
\left( 
\begin{array}{llll}
{1\over\sqrt{N}} & 
 {1\over\sqrt{N}} &  \ldots& {1\over\sqrt{N}} \end{array} \right) =
N {1\over N^2} + N(N-1) {1\over N} a =
\frac{1}{N} + (N-1) a .\]
Since $F(\rho)\geq 1-\epsilon$, it must be the case that
$a\geq {1\over N}-{\epsilon \over N-1}$.
By substituting that into (\ref{eq-andris1}),
the fidelity of the final state with $\ket{\Psi_K}$ is at least
\[ {1\over M} + N\left( {1\over N}-{\epsilon \over N-1} \right)  
\left( 1-{1\over M} \right) 
= 1 - {N\over N-1}\left( 1-{1\over M} \right) \epsilon.  \]

\end{proof}

To prove the second part of Theorem \ref{thm:succ-bound}, 
it remains to show that the protocol 
also succeeds for states not in the diagonal
subspace. 
Let $\ket{\phi}$ be a state such that $F(\ket{\phi})\geq 1-\epsilon$.
We decompose 
\[ \ket{\phi}=\sqrt{1-\delta} \ket{\phi_1}+\sqrt{\delta} \ket{\phi_2},\] 
with $\ket{\phi_1}\in{\cal H}^D$ and $\ket{\phi_2}\in({\cal H}^D)^{\perp}$.
Let $F(\ket{\phi_1})=1-\delta'$.
Since $\Psi_N$ is in ${\cal H}^D$ and $\ket{\phi_2}$ is orthogonal to
${\cal H}^D$, we have $F(\ket{\phi_2})=0$ and 
$F(\ket{\phi})=(1-\delta)(1-\delta')$.
Notice that $(1-\delta)(1-\delta')\geq 1-\epsilon$
because $F(\ket{\psi})\geq 1-\epsilon$.

Applying $U_\pi$ maps $\ket{\phi}$ to 
$\ket{\phi_\pi}=\sqrt{1-\delta} \ket{\phi_{\pi,1}}+\sqrt{\delta} 
\ket{\phi_{\pi,2}}$ where $\ket{\phi_{\pi, 1}}=U_{\pi} \ket{\phi_1}$,
$\ket{\phi_{\pi, 2}}=U_{\pi} \ket{\phi_2}$.
Since $U_{\pi}$ preserves the diagonal subspace,
$\ket{\phi_{\pi, 1}}\in {\cal H^D}$ and
$\ket{\phi_{\pi, 2}}\in ({\cal H^D})^{\perp}$.
Measuring ${\cal H}_L^A$ and ${\cal H}_L^B$
for a state in ${\cal H^D}$ always gives the same
results and produces a state in the diagonal subspace of 
${\cal H}^A_M\otimes {\cal H}^B_M$. 
Measuring ${\cal H}_L^A$ and ${\cal H}_L^B$
for a state in $({\cal H^D})^{\perp}$ either gives the different
results for Alice and Bob or gives the same results but
produces a state orthogonal to the diagonal subspace of 
${\cal H}^A_M\otimes {\cal H}^B_M$. 

The fidelity of the final state consists of two parts:
the fidelity of the final state if Alice's and Bob's measurements
of ${\cal H}_L$ give the same answer and the fidelity if 
measurements give the different answer. 
The first part is just $(1-\delta)$ times the fidelity
of the final state if the starting state was $\ket{\phi_1}$ 
(instead of $\ket{\phi}$). 
Since $\ket{\phi_1}$ is in the diagonal subspace,
Lemma \ref{lem:asp-diagonal} implies that the final state
of the protocol $\ket{\phi_1}$ has the fidelity at least
$1-D\delta'$ where $D={M-1\over M}{N\over N-1}$.
Therefore, the first part is at least
\begin{equation}
\label{eq-andris2} 
(1-\delta)(1-D\delta')= (1-\delta)(1-\delta') + (1-D)\delta'(1-\delta) 
\end{equation}
The second part is the probability of measurements giving
different answers times the fidelity of the state $\ket{0}\otimes\ket{0}$
which Alice and Bob output in this case.
The fidelity of this state is $\frac{1}{M}$ and the
probability of this case is given by

\begin{lemma}
\label{lem-andris1}
The probability that Alice's and Bob's measurements give different
answers is $\frac{N-M}{N-1}\delta$.
\end{lemma}

\begin{proof}
First, we look at the state $\ket{\phi_2}$.
Since this state is in $({\cal H^D})^{\perp}$, it is of the form
\[ \ket{\phi_2} = \sum_{i,j=1, i\neq j}^N \alpha_{i, j} 
\ket{i}^A \ket{j}^B .\]
Applying $U_\pi$ maps it to 
\[ \ket{\phi_{\pi, 2}}= \sum_{i\neq j} \alpha_{i, j} 
\ket{\pi(i)}^A\ket{\pi(j)}^B
= \sum_{i\neq j} \alpha_{\pi^{-1}(i), \pi^{-1}(j)} 
\ket{i}^A\ket{j}^B.\]
The probability of Alice and Bob getting different results is
equal to the sum of $|\alpha_{\pi^{-1}(i), \pi^{-1}(j)}|^2$
over all basis $\ket{i}^A$, $\ket{j}^B$ that differ
in the ${\cal H}_L$ part.
If this sum is averaged over all permutations $\pi$, it
becomes the same for all $i, j$, $i\neq j$.
Therefore, the probability of Alice and Bob getting
different results is just the fraction of pairs $(i, j)$ that
differ in the ${\cal H}_L$ part. 
It is $\frac{N-M}{N-1}$ because for each $i$, there are $(N-1)$ 
$j\in\{1, \ldots, N\}$, $j\neq i$ and $M-1$ of them differ only 
in the ${\cal H}_K$ but the remaining $N-M$ differ in the ${\cal H}_L$ part.

If the starting state is $\ket{\phi}$, the probability of 
Alice and Bob getting different results is $\delta$ times
the probability for $\ket{\phi_2}$ because 
$\ket{\phi}=\sqrt{1-\delta} \ket{\phi_1}+\sqrt{\delta}\ket{\phi_2}$ and 
the measurements always give
the same answer on $\ket{\phi_1}$.
\end{proof}

Therefore, the second part of the fidelity is 
$\frac{1}{M} \frac{N-M}{N-1} \delta$.
Notice that $1-D=1-\frac{(M-1) N}{M(N-1)} = \frac{(M-1) N - M(N-1)}{M(N-1)}=
\frac{N-M}{M(N-1)}$.
Thus, the second part is $(1-D)\delta$ and  
the overall fidelity is at least 
\[
(1-\delta) (1-\delta') + 
(1-D) (1-\delta)\delta' + (1-D) \delta 
= 1 - D(\delta(1-\delta')+\delta') .
\]
Since $(1-\delta)(1-\delta')\geq 1-\epsilon$, 
$\delta (1-\delta')+\delta'\leq \epsilon$.
Therefore, the overall fidelity is at least $1-D\epsilon$.
This completes the proof of the second part of 
Theorem \ref{thm:succ-bound} for $K=1$.

For $K>1$, we can just produce an entangled state of dimension
$M'=M/K$ without the use of $\ket{\Psi_K}$ by the protocol above
and then output this state and the original $\ket{\Psi_K}$.
This achieves the fidelity of at least $1-D\epsilon$
for $D=\frac{M'-1}{M'}\frac{N}{N-1} = \frac{M/K-1}{M/K}\frac{N}{N-1}=
\frac{M-K}{M}\frac{N}{N-1}$, proving that the bound
of Theorem \ref{thm:succ-bound} (a) is tight for $K>1$ as well.


\section{Proof to Lemma~\ref{lemma:negative-simple-bi}}
\label{sec:proof-neg-simple}
We give a (somewhat) simpler proof to
Lemma~\ref{lemma:negative-simple-bi} than the proof by Vidal, Jonathan, and
Nielsen~\cite{VJN00}.

For a self-adjoint matrix $M$, we define its \emph{spectrum}
written as ${\cal S}(M)$, to be a vector formed by
the eigenvalues of $M$, and whose entries are sorted in a
decreasing order. In other words, if the eigenvalues of $M$ are
$\lambda_1, \lambda_2, ..., \lambda_d$, where 
$\lambda_1\ge \lambda_2\ge\cdots\ge\lambda_d$, then 
${\cal S}(M) = (\lambda_1, \lambda_2, ..., \lambda_d)$.

For a mixed state $\rho$, if we write $\rho$ as 
$$\rho = \sum_{i=1}^d{p_i}\cdot \dyad{\phi_i}{\phi_i}$$
where $p_1\ge p_2\ge \cdots \ge p_d$, and $\{\ket{\phi_i}\}$ is an
orthonormal basis, then  
$${\cal S}(\rho) = (p_1, p_2, ..., p_d)$$  

A useful Fact about the spectrum of a tensor product of two matrices
is the following:
\begin{fact}
\label{thm:spectrum}
Let $A$ and $B$ be square matrices such that the eigenvalues for $A$
are $\{\lambda_1, \lambda_2, ..., \lambda_m\}$ and the eigenvalues for
$B$ are $\{\mu_1, \mu_2, ..., \mu_m\}$. Then the eigenvalues for the
matrix $A\otimes B$ are 
$\{\lambda_i\cdot \mu_j\}_{i=1, 2, ..., m,\; j = 1, 2,  .., n}$.
\end{fact}
\begin{proof}
It is easy to verify that if $A\cdot \vec{v} = \lambda\cdot \vec{v}$
and  
$B\cdot \vec{u} = \mu\cdot \vec{u}$, then 
\begin{math}
(A\otimes B)\cdot \vec{(v\otimes u)} = 
(\lambda\cdot\mu)\vec{(v\otimes u)}
\end{math}
\end{proof}

and a corollary the above fact is:
\begin{corollary}
\label{cor:spectrum}
Let $\rho^A$, $\rho^B$ be the density matrices for quantum systems
${\cal H}^A$ and ${\cal H}^B$. Then we have
\begin{equation}
\rank{\rho^A\otimes\rho^B} \ge \rank{\rho^A}
\end{equation}
\end{corollary}
\begin{proof}
Notice that the rank of a matrix equals the number of non-zero
eigenvalues of this matrix. Since $\rho^B$ is a density matrix, it has
trace 1, and thus it has at least one non-zero eigenvalue --- assume
it is $\mu_1$. We denote the eigenvalues of $\rho^A$ by 
$\lambda_1, \lambda_2, ..., \lambda_m$, then by
Fact~\ref{thm:spectrum}, 
$\lambda_1\cdot \mu_1, \lambda_2\cdot \mu_1, ..., \lambda_m\cdot \mu_1$
are all eigenvalues of $\rho^A\cdot\rho^B$, and they contain at many
non-zero numbers as the eigenvalues of $\rho^A$.
\end{proof}

\begin{proof}[Proof to Lemma~\ref{lemma:negative-simple-bi}]

We consider an arbitrary protocol ${\cal P}$ between Alice and Bob
involving only LOCC. We assume that ${\cal P} $ consists of
\emph{steps}, where each step could be one of the following
operations~\footnote{We assume that Alice have enough ancillary qubit
  at the beginning of the protocol and not more new ancillary qubits
  need to be introduced during the protocol.}:
\begin{enumerate}
\item{\bf Unitary Operation:}\\
 Alice (or Bob) applies a unitary operation to her (or his)
  subsystem.
\item{\bf Measurement: }\\
 Alice (or Bob) performs a measurement to her (or his) subsystem.
\item{\bf Tracing Out:} \\
 Alice (or Bob) discards part of her (or his) subsystem, or
  equivalently, traces out part of the subsystem.
\item{\bf Classical Operation:} \\
 Alice (or Bob) sends a (classical) message to the other party.
\end{enumerate}

We first convert this protocol ${\cal P}$ into another protocol 
${\cal  P}'$ in the following way: for each tracing-out operation
Alice (or Bob) performs, we insert a measurement operation right
before the tracing-out, and the measurement is a full measurement of
the subsystem to be traced out. Notice that ${\cal P}'$ will have
exactly the same output as ${\cal P}$, since the subsystem that was
traced out isn't part of the output. However, ${\cal P}'$ has the
property that for each subsystem traced out in the protocol, that
subsystem is disentangled from the rest, since it is already
completely measured.

Now we analyze the new protocol ${\cal P}'$.
We denote the partial density matrix of Alice for the state
$\ket{\phi}$ by $\rho^A$: 
\begin{equation}
\rho^A = \tr_B(\dyad{\phi}{\phi})
\end{equation}

Since we know $\ket{\phi}$ precisely, we can compute $\rho^A$
precisely, and in particular, its spectrum. It is easy to verify that
the spectrum of $\rho^A$ is
$${\cal S}(\rho^A) = (
\underbrace{{1/ K}, {1/ K}, ... {1/K}}_{K}, 
\underbrace{0, 0, ..., 0}_{(N-1)K})
$$
So the rank of $\rho^A$ (which is also the Schmidt Number of
$\ket{\phi}$) is $K$.

We focus on how $\rho^A$ changes with the local operations Alice
performs (apparently it doesn't change with Bob's local
operations): we shall prove that the rank of $\rho^A$ never increases.
 There are 3 types of operations Alice can perform:
unitary operations, local measurements, and tracing out a
subsystem, we analyze them one by one:
\begin{itemize}
\item {\bf Unitary Operations}\\
This operation changes a mixed state $\rho^A$ to 
$U\rho^A U^\dagger$, where $U$ is a unitary operation. Obviously the
rank doesn't change.
\item {\bf Local Measurements} \\
Suppose measurement operator is $\{M_m\}$ satisfying
$\sum_{m}M_m^\dagger M_m = I$, and the measurement yields
result $m$. Then Alice ends in state 
$$\rho_m = {{M_m\rho^AM_m^\dagger}\over{\tr(M_m^\dagger M_m\rho^A)}}
$$
Again, we have $\rank{\rho_m}\le\rank{\rho^A}$.
\item {\bf Tracing Out a Subsystem}\\
We write ${\cal H}^A = {\cal H}^{A_0} \otimes {\cal H}^{A_1}$, and we
suppose that the subsystem ${\cal H}^{A_1}$ is traced out. We write
the partial density matrix for ${\cal H}^{A_0}$ as $\rho^{A_0}$, and
we have $\rho^{A_0} = \tr_{A_1}(\rho^A)$.

We know that in protocol ${\cal P}'$, the subsystem ${\cal H}^{A_0}$
is disentangled from the subsystem ${\cal H}^{A_1}$. Thus we have

$$\rho^A = \rho^{A_0}\otimes\rho^{A_1}$$
for some density matrix $\rho^{A_1}$.
and by Corollary~\ref{cor:spectrum}, we have
$\rank{\rho^{A_0}}\le \rank{\rho^A}$.

\end{itemize}
So, as Alice and Bob perform local operations, the rank of the partial
density matrix for Alice never increases. This fact remains true even
if Alice and Bob perform classical communications (this just means
that Alice has the ability to perform different local operations
according to Bob's measurement result, but no local operation Alice
performs can increase the rank).

We denote the density matrix for the final state after the protocol
${\cal P}$ to be $\rho_E$, and we define 
$\rho^A_E = \tr_B(\rho_E)$ to be the partial density matrix for
Alice. Then we have $\rank{\rho^A_E}\le K$. Notice $\rho^A_E$ should
be an $M\times M$ matrix since Alice and Bob 
are supposed to arrive at a state in 
${\cal H}_{M}^A\otimes{\cal H}_{M}^B$. 
We use $\rho_0^A$ to denote the partial density matrix for Alice if we
trace out the system ${\cal H}^B_{M}$ from the target state 
$\Psi_{M}$. It
is easy to verify that $\rho_0^A = {1\over M}I$, where $I$ is the
identity matrix. 

By monotonicity of fidelity, we have
$$ F(\rho_E, \dyad{\Psi_{M}}{\Psi_{M}}) \le F(\rho_E^A, \rho_0^A)$$
However, we have
\begin{eqnarray*}
F(\rho_E^A, \rho_0^A) & = & \tr\sqrt{(\rho_E^A)^{1/2}\rho_0^A
(\rho_E^A)^{1/2}} \\
& = & \sqrt{1\over M}\cdot \tr\sqrt{\rho^A_E}
\end{eqnarray*}
We write the spectrum of $\rho^A_E$ as 
$${\cal S}(\rho^A_E) = (\lambda_1, \lambda_2, ..., \lambda_{M})$$
and we know that 
$\lambda_{K+1} = \lambda_{K+2} = \cdots = \lambda_{M} = 0$ since 
$\rank{\rho^A_E} \le K$.
Therefore, we have
\begin{eqnarray*}
\tr\sqrt{\rho^A_E} & = & \sum_{l=1}^{M}\sqrt{\lambda_l}
= \sum_{l=1}^{K}\sqrt{\lambda_l} 
\le \sqrt{K}\cdot \left(\sum_{l=1}^{K}{\lambda_l}\right) 
 = \sqrt{K}
\end{eqnarray*}
and thus
$$F(\rho_E^A, \rho_0^A) = \sqrt{1\over M}\cdot \tr\sqrt{\rho^A_E} 
\le \sqrt{K\over M}$$
Therefore we have
$$F(\rho_E) = F(\rho_E, \dyad{\Psi_{M}}{\Psi_{M}}) \le F(\rho_E^A,
\rho_0^A) \le {K\over M}$$
\end{proof}


\section{Constructions of Deterministically Conditionally Successful
  GEPPs} 
\label{appendix-2}

We describe the construction of the ``Simple Scrambling'' protocol.
This protocol is deterministically conditionally successful GEPP for
input states in the diagonal subspace. 

The construction relies on a
special family of permutations, namely the Scrambling Permutations. We
give a definition of Scrambling Permutations first, and postpone the
actual construction of these permutations to the
Appendix~\ref{sec:construct-scramble}.

\subsection{Scrambling Permutations}
We define a class of permutations that would be useful in constructing
the Simple Scrambling protocol.
We work on functions over binary strings, and we use $x\circ y$
to denote string $x$ concatenated with string $y$.
For finite sets $A$
and $B$ of binary strings, we define the \emph{concatenation} of $A$
and $B$ to be set
$$A\circ B = \{a\circ b\;|\;a\in A, b\in B\}$$

We will be working with 4 finite sets of binary strings, and we call
them $X$, $Y$, $G$, and $H$. These sets have the property that $X =
G\circ H$. We define $N= |X|$, $K=|Y|$, $W=|H|$, 
$L=|G|$, and we will be using this size convention for the rest of
this paper. Obviously we have $N = WL$.

\begin{definition}[Scrambling Permutation]
\label{def:scramble}
A class of parameterized function pairs
$\langle g_y(x), h_y(x) \rangle$ of types $g_y:X\mapsto G$ and 
$h_y:X\mapsto H$ is called a {\em scrambling  permutation pair of
  parameter $(N, K, W, L)$}, or simply \emph{scrambling permutation},
if the following 2 conditions are satisfied:
\begin{enumerate}
\item {\bf (Permutation)}
  For all $y\in Y$, $x\longmapsto g_y(x)\circ h_y(x)$
  is a permutation in $X$.
\item{\bf (Scrambling)}
  There exists a positive number $p$, such that 
  or any pair of elements $x_1\ne x_2$ in $X$, 
  $$\prob_y[h_y(x_1) = h_y(x_2)] = p$$
  where the probability is taken over the $y$ uniformly chosen from
  $Y$. We call this $p$ the ``collision probability''.
\end{enumerate}
Furthermore, the pair will be called \emph{efficient scrambling
  permutation pair} if both the function $g_y(x)\circ h_y(x)$ and its
inverse can be efficiently computed (i.e., has polynomial-size circuits).
\end{definition}

It is interesting to compare the definition of scrambling
permutations to that of universal hash functions~\cite{CW79, WC81}.
On one hand, the 
scrambling permutations are permutations, while the universal hash
functions don't have to be. On the other hand, the ``scrambling''
property in the scrambling permutation is weaker than that of the
universal hash functions. For scrambling permutations, the function
$h_y(x)$ only need to have a constant collision probability for
all pairs $(x_1, x_2)$. However, for universal hash functions, 
the setting is that $\prob_y[h_y(x_1) =a \;\land\;h_y(x_2)=b]$ is the
same for all $(x_1, x_2, a, b)$ tuples. Obviously, any universal hash
function that can be extended to a permutation will induce a
scrambling permutation. An example is the linear map construction 
($f_{a,b}(x) = a\cdot x + b$, see~\cite{MR95}, page 219,
or~\cite{L96}, page 85). However, there exist more efficient
constructions of scrambling permutations --- We postpone the detailed
construction and discussion to Appendix~\ref{sec:construct-scramble},
and we just state the results here:

\begin{theorem}
\label{thm:construct-scramble}
There exist efficient scrambling permutations with the following
parameters:
\begin{itemize}
\item $\langle 2^n, (2^n-1), 2^{n-t}, 2^t\rangle$
\item $\langle 2^{2n}, (2^n+1)2^{2t}, 2^{n}, 2^n\rangle$
\item 
\begin{math}
\langle 2^{dn}, {{2^{dn}-1}\over{2^n-1}}\cdot 2^{2t},
2^{(d-1)n}, 2^n\rangle
\end{math} 
\end{itemize}
where $n,t,d$ are integers such that $n>t$.
\qed
\end{theorem}

A useful fact about Scrambling Permutation is that the collision
probability $p$ can be computed.

\begin{theorem}
\label{thm:scramble-perm-collision}
Let $\langle g_y(x), h_y(x)\rangle$  be a
scrambling permutation pairs of parameter $\langle N, K, W, L\rangle$.
The collision probability $p$ equals $(L-1)/(N-1)$.
\end{theorem}

\begin{proof}

We call a triple $\langle x_1, x_2, y\rangle$ a ``collision
instance'', if  $h_y(x_1) = h_y(x_2)$. Now we count how many such
collision instances there are. There are two ways to count 
them. 
\begin{itemize}
\item For each $(x_1, x_2)$ pair, there are $K\cdot p$ $y$'s such that
  $h_y(x_1) = h_y(x_2)$. So the total number of collision instances is
  $$Kp\cdot N(N-1)/2$$
\item For each fixed $h_y(\cdot)$, it is a function that maps $X$ of
  size $N$ to $H$ of size $W$. Since $g_y\cdot h_y$ is a permutation,
  the mapping $h_y(x)$ has to be an ``even'' one: for each $u\in H$,
  there must be precisely $L$ elements in $X$ that are mapped to $u$.
  So the $N$ elements in $X$ are partitioned into $W$ subsets, each of
  size $L$. The number of pairs that are in the same subset is therefore
  $W\cdot L(L-1)/2$.  So the number of collision instances is
  $$K\cdot W \cdot L(L-1)/2$$
\end{itemize}
The two ways should give the same result. Thus we have
$$Kp\cdot N(N-1)/2 = K\cdot W \cdot L(L-1)/2$$
or
$$p = {{L-1}\over{N-1}}$$

\end{proof}

\subsection{The Construction of the Simple Scrambling Protocol}

We describe the construction of the Simple Scrambling
protocol, which is
deterministically conditional successful for input states in the
diagonal subspace.

First, we recall the definitions of the Fourier Operator and the
Hadamard Operator. 

For a Hilbert space of dimension $N$, the \emph{Fourier
Operator} is defined by the matrix $F$ where
the $(x,y)$-th entry of $F$ is ${1\over\sqrt{N}}\omega^{-x\cdot y}$,
for $x,y = 0, 1, ..., N-1$, where  $\omega = e^{i{2\pi\over N}}$ is a
root of the unity, and $x\cdot y$ denotes the integer
multiplication. $F$ is a unitary operator. We call its 
inverse, $F^\dagger$, the \emph{Inverse Fourier Operator}.

For a Hilbert space of dimension $N=2^n$, the \emph{Hadamard
    Operator} is defined by the matrix $H$ where the
$(x,y)$-th entry of $H$ is ${1\over\sqrt{N}}(-1)^{x \bullet y}$, where
$x\bullet y$ is the inner product of $x$ and $y$, for 
$x, y = 0, 1,..., N-1$.  

This protocol is parameterized by 4 integers: $N, K, W, L$, such that
there exists a scrambling permutation pair 
$\langle g_y(x), h_y(x)\rangle$ of parameter $(N, K, W, L)$.

\begin{construction}[Simple Scrambling Protocol]
\label{construct:simple-scramble}
The input to the protocol is a state $\rho$ in 
${\cal  H}_N^A\otimes{\cal H}_N^B$. The protocol also
has an auxiliary input of $\Psi_K$. The steps are:
\begin{enumerate}
\item Alice and Bob both apply the scrambling permutation to their
  qubits: using the qubits from $\rho$ as $x$ and the qubits
  from 
  $\Phi_K$ as $y$, and outputs the values of both functions and $y$:
  \begin{equation}
    \label{eqn:mapping}
    \ket{x}\ket{y} \;\longrightarrow
    \ket{g_y(x)}\ket{h_y(x)}\ket{y}
  \end{equation}
  Here we identify the Hilbert space ${\cal H}_N\otimes{\cal H}_K$
  with ${\cal H}_L\otimes{\cal H}_W\otimes{\cal H}_K$.

\item Alice applies the  Fourier Operator to the state 
$\ket{g_y(x)}^A$, and Bob applies the Inverse Fourier Operator to the
state $\ket{g_y(x)}^B$. Then both measure these qubits in the
computational basis.

In the case that $L$ is a power of 2, Alice and Bob can, alternatively,
both apply a Hadamard operator to their states of $\ket{g_y(x)}$,
instead of the Fourier and Inverse Fourier operators.

\item Alice and Bob compare their results via classical communication.
\item
  If the results are the same, they discard the measured state (or
  equivalently, trace out the subspace ${\cal H}_L$), and output the
  remaining state, which is in Hilbert space 
  ${\cal H}_{WK}^A\otimes{\cal H}_{WK}^B$.
\item
  If the results are different, they discard everything and output
  $\fail$. 
\end{enumerate}

\end{construction}

We point out that this protocol can be implemented by LOCC. In step 1,
both Alice and Bob apply a scrambling permutation to their state.
It is easy to
verify that the mapping in Equation~\ref{eqn:mapping} is a permutation
and thus is possible to realize quantum-mechanically. Next, if the
scrambling permutation is efficient, there exists a polynomial-size
quantum circuit that implements it~\cite{L01}. In step 2, Fourier
Operators and Inverse Fourier Operators are applied by Alice and
Bob, respectively. Fourier Operators exist for every $N$ and when $N$
is a power of 2, there exists an efficient implementation of both the
Fourier Operators and Inverse Fourier Operators. Also, in the case $N$
is a power of 2, there exists a very efficient algorithm for performing
Hadamard operators. 
Therefore we have:
\begin{claim}
\label{claim:implement-simple}
The simple scrambling protocol can be implemented by LOCC.
Furthermore, if the scrambling permutation used
in the protocol is efficient and $L$ is power of 2, the protocol can
be efficiently implemented.
\end{claim}

\subsection{The Analysis of the Simple Scrambling Protocol}

Now we prove a lemma that the Simple Scrambling protocol is
deterministically conditionally successful for input states that are
\emph{pure states} in the diagonal subspace.

\begin{lemma}
\label{lemma:simplified-fidelity}
If the input state to a Simple Scrambling protocol is a pure state in
the diagonal subspace,
then this protocol is deterministically conditionally successful with
parameter 
$\langle N, K, WK, \epsilon, {2W\over{N}}\epsilon, \epsilon\rangle$
for $\epsilon < 1/2$.
\end{lemma}

Intuitively, this lemma is true because the Scrambling Permutation
``shuffles'' the coefficients of $\ket{\phi}$ very
``evenly''. Then the Fourier operator (or the Hadamard operator)
``mixes'' all the coefficients together. Therefore when the
protocol doesn't fail, the coefficients of the output state are much
more ``smooth'' than that of $\ket{\phi}$.

\begin{proof}
We write the input state $\ket{\phi}$ as
\begin{equation}
\ket{\phi} = \sum_{x\in X}\alpha_x\ket{x}^A\ket{x}^B
\end{equation}
and we have that 
$$\sum_{x\in X}|\alpha_x|^2 = 1$$
We denote $\sum_{x\in X}\alpha_x$ by $D$. Then we have

\begin{equation}
1-\epsilon = |\braket{\phi}{\Psi_N}|^2 
 =  {1\over N}\cdot{\left|\sum_{x\in X}\alpha_x\right|^2} =
 {{|D|^2}\over N} 
\end{equation}

We will go through the protocol and keep track of the state.

\begin{enumerate}
\item The initial state for Alice and Bob is
\begin{equation}
\ket{\psi_1}  =  \ket{\phi}\otimes \Psi_K = 
{1\over{\sqrt{K}}}\sum_{x\in X}\sum_{y\in Y}\alpha_x \ket{x\circ y}^A
\ket{x\circ y}^B
\end{equation}
\item
After applying the scrambling permutation, the state becomes
\begin{equation}
\ket{\psi_2} = 
{1\over{\sqrt{K}}}\sum_{x\in X}\sum_{y\in Y}\alpha_x \cdot
\ket{g_y(x)\circ h_y(x)\circ y}^A\cdot\ket{g_y(x)\circ h_y(x)\circ y}^B
\end{equation}
\item
After the Fourier and Inverse Fourier operators, the state is
\begin{equation}
\ket{\psi_3} = {1\over{L\sqrt{K}}}\sum_{x\in X}
\sum_{y\in Y}\sum_{g_A\in G}\sum_{g_B\in G}
\omega^{g_y(x)\cdot(g_B - g_A)}\cdot\alpha_x\cdot
\ket{g_A\circ h_y(x)\circ y}^A\cdot\ket{g_B\circ h_y(x)\circ y}^B
\end{equation}
Alternatively, if $L$ is a power of 2, and Hadamard operators are used
instead of Fourier and Inverse Fourier operators, 
the state is
\begin{equation}
\ket{\psi'_3} = {1\over{L\sqrt{K}}}\sum_{x\in X}
\sum_{y\in Y}\sum_{g_A\in G}\sum_{g_B\in G}
(-1)^{g_y(x)\bullet(g_A\oplus g_B)}\cdot\alpha_x\cdot
\ket{g_A\circ h_y(x)\circ y}^A\cdot\ket{g_B\circ h_y(x)\circ y}^B
\end{equation}

In either cases, if Alice and Bob both measure their qubits and they
both obtain the result $g$, the state becomes 
\begin{equation}
\ket{\psi_{4, g}} = {\Delta\over{L\sqrt{K}}}\sum_{x\in X}
\sum_{y\in Y}\alpha_x\cdot
\ket{g\circ h_y(x)\circ y}^A\cdot\ket{g\circ h_y(x)\circ y}^B
\end{equation}
where $\Delta$ is a normalization factor. Notice that if Alice and Bob
both discard the qubits $\ket{g}^A$ and $\ket{g}^B$  (which are
disentangled from the rest), the resultant state is the same for
different $\ket{g}$'s:
\begin{equation}
\ket{\psi_5} = {\Delta\over{L\sqrt{K}}}\sum_{x\in X}
\sum_{y\in Y}\alpha_x\cdot
\ket{h_y(x)\circ y}^A\cdot\ket{h_y(x)\circ y}^B
\end{equation}
\end{enumerate}
Now let's compute $\Delta$.

To normalize $\ket{\psi_5}$, we can re-group the terms and re-write it
as 
\begin{equation}
\ket{\psi_5} = {\Delta\over{L\sqrt{K}}}\sum_{u\in H}
\sum_{y\in Y}\left(
\sum_{h_y(x)=u} \alpha_x
\right)\cdot
\ket{u\circ y}^A\cdot\ket{u\circ y}^B
\end{equation}
Therefore we should have

\begin{eqnarray*}
{{\Delta^2}\over{L^2K}}\sum_{u\in H}
\sum_{y\in Y}\left|\sum_{h_y(x)=u} \alpha_x\right|^2 =1 
\end{eqnarray*}

Furthermore, we have

\begin{eqnarray*}
\sum_{u\in H}\sum_{y\in Y}\left|\sum_{h_y(x)=u} \alpha_x\right|^2
& = & 
\sum_{y\in Y}\sum_{u\in H}\left|\sum_{h_y(x)=u} \alpha_x\right|^2\\
& = & 
\sum_{y\in Y}\sum_{x\in X}|\alpha_x|^2 +
\sum_{y\in Y}\;\sum_{x_1\ne x_2, h_y(x_1) = h_y(x_2)}
(\alpha_{x_1}\overline{\alpha}_{x_2} + 
\overline{\alpha}_{x_1}\alpha_{x_2})\\
& = & \sum_{y\in Y}1 +
\sum_{x_1\ne x_2}\sum_{y\in Y: h_y(x_1) = h_y(x_2)}
(\alpha_{x_1}\bar{\alpha}_{x_2} 
+\bar{\alpha}_{x_1}\alpha_{x_2})\\
& = & K + \sum_{x_1\ne x_2}pK (\alpha_{x_1}\bar{\alpha}_{x_2}
+\bar{\alpha}_{x_1}\alpha_{x_2})\\
& = & K + pK\left(\left|\sum_{x\in X}\alpha_x\right|^2
- \sum_{x\in X}|\alpha_x|^2 \right) \\
& = & K[1 + p(|D|^2-1)]
\end{eqnarray*}

and thus
\begin{eqnarray*}
\Delta^2 & = & {{L^2K}\over{ K[1 + p(|D|^2-1)]}} \\
& = & {{L^2}\over{1 + p(|D|^2-1)}}\\
& = & {{L^2}\over{1 + {{L-1}\over{N-1}}(N-1-N\epsilon)}}\\
& = & {L\over{1-\epsilon\cdot{{N(L-1)}\over{L(N-1)}}}}
\end{eqnarray*}

Notice $\Delta^2$ is the probability that Alice and Bob both obtain
$\ket{g}$ for their measurement. There are $L$ possible $\ket{g}$'s
that Alice and Bob can obtain.
So the probability that Alice and Bob obtain the same result is
$$\prob[\mbox{\sf Alice and Bob obtain the same result}]
= {L\over{\Delta^2}} = 1-\epsilon\cdot{{N(L-1)}\over{L(N-1)}} \ge
1-\epsilon$$
And the fidelity of $\ket{\psi_5}$ is
\begin{eqnarray*}
F(\dyad{\psi_5}{\psi_5}) & = & 
|\braket{\psi_5}{\Psi_{WK}}|^2 \\
& = & 
\left|
{1\over{\sqrt{WK}}}\cdot {\Delta\over{L\sqrt{K}}}\sum_{u\in H}
\sum_{y\in Y}\left(
\sum_{h_y(x)=u} \alpha_x
\right) \right|^2\\
& = & 
{{\Delta^2}\over{L^2K^2W}}
\left|\sum_{u\in H}\sum_{y\in Y}\sum_{h_y(x)=u}
\alpha_x \right|^2\\
& = & {{\Delta^2}\over{L^2K^2W}}\left|
\sum_{x\in X}\sum_{y\in Y} \alpha_x \right|^2\\
& = & {{\Delta^2}\over{L^2K^2W}}\cdot |KD|^2 \\
& = & {{(1-\epsilon)\cdot \Delta^2}\over L} \\
& = & {{1-\epsilon}\over{1-\epsilon\cdot{{N(L-1)}\over{L(N-1)}}}} \\
& = & 1 - \epsilon\cdot {{N-L}\over{L(N-1)}}\cdot 
{1\over{1-\epsilon\cdot{{N(L-1)}\over{L(N-1)}}}} \\
& \ge & 1-2\epsilon\cdot  {{N-L}\over{L(N-1)}} \\
& = & 1-2\epsilon\cdot {{W-1}\over{N-1}} \\
& \ge & 1 - {2W\over N}\epsilon
\end{eqnarray*}

\end{proof}

Therefore, if the input state $\ket{\phi}$ has a sufficiently
high fidelity, then with high probability the Simple Scrambling
protocol will succeed, and the resultant state, which is also a pure
state, can have a much higher fidelity.

Next, we prove that the Simple Scrambling protocol
works without modification for mixed states in Diagonal
Subspaces, with exactly the same parameters.

\begin{lemma}
\label{lemma:mixed-fidelity}
If the input state to a Simple Scrambling protocol is a mixed state in
the diagonal subspace, then this protocol is deterministically
conditionally successful with parameter 
$\langle N, K, WK, \epsilon, {2W\over{N}}\epsilon, \epsilon\rangle$
for $\epsilon < 1/2$.
\end{lemma}
\begin{proof}
We write the mixed state as an ensemble: 
$\{p_i,\ket{\phi_i}\}_{i=1,2,... s}$. We use $1-\epsilon_i$ to denote
the fidelity of pure state $\phi_i$. Then we have 
$\epsilon = \sum_{i}p_i\cdot \epsilon_i$ by the linearity of fidelity.

As in the proof to Lemma~\ref{lemma:simplified-fidelity}, the protocol
will succeed with probability
$1-\epsilon_i\cdot{{N(L-1)}\over{L(N-1)}}$ for state
$\ket{\phi_i}$. So the overall probability that the protocol doesn't
fail is
\begin{eqnarray*}
\prob[{\cal P}(\dyad{\phi_i}{\phi_i}) = \fail]
& = & \sum_{i}p_i\cdot
\epsilon_i\cdot{{N(L-1)}\over{L(N-1)}}\\ 
& = & {{N(L-1)}\over{L(N-1)}}\cdot\sum_{i}p_i\cdot\epsilon_i \\
& = & {{N(L-1)}\over{L(N-1)}}\cdot \epsilon \\
& \le & \epsilon
\end{eqnarray*}

If we use $Q_i$ to denote the fidelity of the output of the protocol on
state $\ket{\phi_i}$ conditioned on that it doesn't fail, then the
overall fidelity of the output if the protocol doesn't fail is:
\begin{eqnarray*}
Q & = & {{\sum_{i}Q_i\cdot p_i\cdot \prob[
{\cal P}(\dyad{\phi_i}{\phi_i})=\fail]}
\over
{\sum_{i}p_i\cdot \prob[{\cal P}(\dyad{\phi_i}{\phi_i})=\fail]}} \\
& = & {{\sum_i {{1-\epsilon_i}\over{1-\epsilon_i\cdot{{N(L-1)}
\over{L(N-1)}}}}\cdot
p_i\cdot \left[1-\epsilon_i\cdot{{N(L-1)}\over{L(N-1)}}\right]
}
\over{\sum_i
p_i\cdot \left[1-\epsilon_i\cdot{{N(L-1)}\over{L(N-1)}}\right]
}}\\
& = & {{\sum_i(1-\epsilon_i)\cdot p_i}\over
{\sum_i(1-{{N(L-1)}\over{L(N-1)}}\cdot\epsilon_i)\cdot p_i}}\\
& = & {{1-\epsilon}\over{1-\epsilon\cdot{{N(L-1)}\over{L(N-1)}}}} \\
& \ge & 1-{2W\over N}\epsilon
\end{eqnarray*}
So the Simple Scrambling protocol is deterministically conditionally
successful even for mixed states in the diagonal subspace.

\end{proof}

Now we are ready to prove Theorem~\ref{thm:succ-det}.
\begin{proof}[Proof to Theorem~\ref{thm:succ-det}]
We simply combine Theorem~\ref{thm:construct-scramble} and
Lemma~\ref{lemma:mixed-fidelity}.
\end{proof}

\section{Toward the Diagonal Subspace: The Hash and Compare Protocol}
\label{appendix-3}

In this section, we present the construction of the Hash and
Compare protocol. With high probability, this protocol converts any
state of reasonably high  fidelity into a state that is ``almost
completely'' in the diagonal subspace. Therefore, if we combine this
protocol with the Simple Scrambling protocol, we obtain a GEPP that is
probabilistically conditionally successful for arbitrary states. 

Before describing the actual protocol, we give some motivations and
intuitions behind it. Suppose Alice and Bob share a state of
fidelity at least $1-\epsilon$. For simplicity, we assume that the
state is a pure state, and we will show how to extend our result to
mixed states later.  We write the input state  as $\ket{\phi}$.
 In this situation, the Simple Scrambling protocol doesn't
work anymore. Essentially what this protocol does is to
``shuffle'' and ``mix'' the coefficients in the diagonal subspace in a
very ``even'' way  to increase the fidelity. The Scrambling
Permutation guarantees that coefficients in the diagonal subspace
will be mixed ``evenly''. However it gives no guarantee for
coefficients outside this subspace. Nevertheless, it is worth noting
that the maximally entangled state, $\Psi_N$, is 
completely in the diagonal subspace. So if $\ket{\phi}$ is close to
$\Psi_N$, then a large ``fraction'' of $\ket{\phi}$ must lie in the
diagonal subspace.

We write 
\begin{equation}
\label{eqn:diagonal-break-down}
\ket{\phi} = \alpha\cdot \ket{\phi_{\parallel}} +
\beta\cdot\ket{\phi_{\perp}}
\end{equation}
where $\ket{\phi_{\parallel}}$ is a vector in the diagonal subspace
${\cal H}^{\cal D}$ and $\ket{\phi_\perp}$ is a vector orthogonal to 
${\cal H}^{\cal D}$ subspace. Both vectors are normalized, and thus we have 
$|\alpha|^2 + |\beta|^2 = 1$. Obviously we have
$\braket{\phi_\perp}{\Psi_N} = 0$ and thus
$|\alpha|^2\ge 1-\epsilon$.

The Simple Scrambling protocol works well for state
$\ket{\phi_\parallel}$, but does not work for state
$\ket{\phi_\perp}$. So if we can first ``eliminate'' 
$\ket{\phi_\perp}$, or at least decrease its coefficient from $\beta$
to a much smaller one, we can use
the Simple Scrambling protocol to obtain a state with high fidelity. 
The Hash and Compare protocol does exactly this.

\begin{construction}[Hash and Compare]
The input to the protocol is a state $\rho$ in the subspace
${\cal H}_N^A\otimes {\cal H}_N^B$.
The protocol also has an auxiliary
input $\Psi_S$,  where $S=2^{s}$ is a power of 2.
The output of the protocol is a 
state $\sigma$ in ${\cal H}_N^A\otimes {\cal H}_N^B$.
The steps are:
\begin{enumerate}
\item 
Alice randomly generates $s$ numbers $r_0, r_1, ..., r_{s-1}\in [N]$
and introduces $s$ ancillary qubits, 
$\ket{b_0}, \ket{b_1}, ..., \ket{b_{s-1}}$,
all initialized to $\ket{0}$. 
\item
Alice performs $s$ unitary operations:
$$\ket{x}\ket{y_j} \longrightarrow \ket{x}\ket{y_j\oplus 
(x\bullet r_j)}$$
She uses the qubits from state $\rho$ as $x$, and the ancillary
qubit $\ket{b_j}$ as $y_j$, for $j=0,1, ..., s-1$.
\item 
Alice send $r_0, r_1, ..., r_{s-1}$ to Bob.
\item 
Alice and Bob engage in $s$ teleportation protocols. They use the
shared state $\Psi_S$ as $s$ EPR pairs, to teleport the $s$ ancillary
qubits $\ket{b_0}^A, \ket{b_1}^A, ..., \ket{b_{s-1}}^A$ from Alice to
Bob. Then Alice discards all her ancillary qubits. Bob obtains the
qubits $\ket{b_0}^B, \ket{b_1}^B, ..., \ket{b_{s-1}}^B$.
\item 
Bob performs $s$ unitary operations (the same operations as Alice did):
$$\ket{x}\ket{y_j} \longrightarrow \ket{x}\ket{y_j\oplus 
(x\bullet r_j)}$$
He uses the qubits from state $\ket{\phi}$ as $x$, and qubit
$\ket{b_j}^B $ as $y_j$, for $j=0, 1, ..., s-1$.
\item 
Bob measures all his ancillary bits 
$\ket{b_0}^B, \ket{b_1}^B, ...,\ket{b_{s-1}}^B$.
\item 
If all the results of the measurements are 0, Bob discards all the
ancillary qubits. Then Alice and Bob output the remaining state,
which is in Hilbert space ${\cal H}_N^A\otimes{\cal H}_N^B$.
\item 
If not all the results of the measurements are 0, Alice and Bob discard
everything and output $\fail$.
\end{enumerate}
\end{construction}

We point out that the Hash and Compare protocol can be
efficiently implemented.

Now we prove that the Hash and Compare protocol will bring the input
state $\ket{\phi}$ to another state that is ``almost'' in the diagonal
subspace. 

First, we extend the definition of fidelity.  We define the fidelity
between a pure state $\ket{\varphi}$ and a linear subspace $L$ to be
the square of the length of the the projection of $\ket{\varphi}$ on
$L$. Alternatively, we have
\begin{equation}
F(\ket{\varphi}, L) = \max_{\ket{\psi}\in L}|\braket{\varphi}{\psi}|^2
\end{equation}
Now we state and prove our lemma about the Hash and Compare protocol.
\begin{lemma}
\label{lemma:hash-compare}
Let state $\ket{\phi}$ be a pure state of fidelity at least
$1-\epsilon$, where $\epsilon < 1/2$.
If $\ket{\phi}$ is the input state to the Hash and Compare protocols,
then the probability this protocol outputs $\fail$ is at most
$\epsilon$. Given that the protocol doesn't fail, we use $\ket{\psi}$
to denote the output state, which is a pure state. We have 
$F(\dyad{\psi}{\psi}) \ge 1-\epsilon$, and 
$$
\prob[F(\ket{\psi}, {\cal H}^{\cal D}) \ge 
1-{{2}\over\sqrt{S}}\epsilon\;]\ge 1-{1\over\sqrt{S}}
$$
\end{lemma}

\begin{proof}
We write the state $\ket{\phi}$ as
\begin{equation}
\ket{\phi} = \sum_{x_A\in X}\sum_{x_B\in X}\alpha_{x_A,x_B}
\ket{x_A}^A\ket{x_B}^B
\end{equation}
and we have
$$\sum_{x_A\in X}\sum_{x_B\in X}|\alpha_{x_A,x_B}|^2 = 1$$
Comparing this to Equation~\ref{eqn:diagonal-break-down}, we conclude
that
\begin{eqnarray*}
|\alpha|^2 & = & \sum_{x\in X}|\alpha_{x,x}|^2 \\
|\beta|^2  & = & \sum_{x_A\ne x_B}|\alpha_{x_A, x_B}|^2 \\
\end{eqnarray*}

We go through the protocol:
\begin{enumerate}
\item
The initial state for Alice and Bob, excluding the auxiliary input
$\Psi_T$ is: 
$$
\ket{\phi_1}  = \sum_{x_A\in X}\sum_{x_B\in X}\alpha_{x_A,x_B}
\ket{x_A}^A\ket{x_B}^B
$$
\item 
After Alice introduces her ancillary qubits and done with the $t$
unitary operations, the state is:
\begin{equation}
\ket{\phi_2} = \sum_{x_A\in X}\sum_{x_B\in X}\alpha_{x_A,x_B}
\ket{x_A}^A\ket{x_A\bullet r_0}^A\ket{x_A\bullet r_1}^A\cdots 
\ket{x_A\bullet r_{s-1}}^A\ket{x_B}^B
\end{equation}
as we can see, the ancillary qubits are entangled with the qubits from
$\ket{\phi}$.
\item 
After the teleportation, Alice's ancillary qubits becomes disentangled
from the qubits of $\ket{\phi}$, and after discarding  all the
ancillary qubits of Alice, the state becomes
\begin{equation}
\ket{\phi_3} = \sum_{x_A\in X}\sum_{x_B\in X}\alpha_{x_A,x_B}
\ket{x_A}^A\ket{x_B}^B\ket{x_A\bullet r_0}^B\ket{x_A\bullet
  r_1}^B\cdots\ket{x_A\bullet r_{s-1}}^B
\end{equation} 
\item 
After Bob has done with his unitary operations, the state becomes
\begin{equation}
\label{eqn:hash-compare-4}
\ket{\phi_4} = \sum_{x_A\in X}\sum_{x_B\in X}\alpha_{x_A,x_B}
\ket{x_A}^A\ket{x_B}^B\ket{(x_A\oplus x_B) \bullet r_0}^B
\ket{(x_A\oplus x_B)\bullet  r_1}^B\cdots\ket{(x_A\oplus x_B)\bullet
  r_{s-1}}^B 
\end{equation}
\item
Next, Bob measures all his ancillary qubits. Now it should be
clear that if the state Alice and Bob start with, $\ket{\phi}$, is
indeed in the diagonal subspace, then all the measurements will yield
0 with probability one, since we have $x_A = x_B$ for all non-zero
$\alpha_{x_A, x_B}$'s.

Now that $\ket{\phi}$ is not in the diagonal subspace, but it is
close. Thus intuitively, Bob should have a high probability getting
all 0's in his measurement.

We do a more formal analysis: we denote by $Z$ the subset of $[N]$
whose elements have inner product 0 with all 
$r_0, r_1, ..., r_{s-1}$: 
$$ Z = \{x\;|\;x\in [N],\; x\bullet r_j = 0, j = 0, 1, ..., s-1\}$$

We group all the terms in Equation~\ref{eqn:hash-compare-4} into 3
parts: 

$$\ket{\phi_4} = \lambda_0 \cdot\ket{\psi_{0}}
+ \lambda_1 \cdot \ket{\psi_{1}} +
\lambda_2 \cdot \ket{\psi_{2}}$$

where 
\begin{eqnarray*}
\lambda_0 \cdot\ket{\psi_0} & = & \sum_{x\in X}
\alpha_{x,x}\cdot \ket{x}^A\ket{x}^B\ket{0}^B\cdots\ket{0}^B\\
\lambda_1 \cdot \ket{\psi_1} & = & \sum_{x_A \ne x_B, x_A\oplus
  x_B\in Z}\alpha_{x_A,x_B}\cdot 
\ket{x_A}^A\ket{x_B}^B\ket{0}^B\cdots\ket{0}^B\\
\lambda_2 \cdot \ket{\psi_2} & = & \sum_{x_A \ne x_B, x_A\oplus
  x_B\not\in Z}\alpha_{x_A,x_B}\cdot 
\ket{x_A}^A\ket{x_B}^B\ket{(x_A\oplus x_B) \bullet r_0}^B
\ket{(x_A\oplus x_B)\bullet  r_1}^B\cdots\ket{(x_A\oplus x_B)\bullet
  r_{s-1}}^B 
\end{eqnarray*}

Both $\ket{\psi_0}$ and $\ket{\psi_1}$ have all 0's in the ancillary
qubits of Bob, while $\ket{\psi_2}$ doesn't. All these 3 states,
$\ket{\psi_0}, \ket{\psi_1}$ and $\ket{\psi_2}$ are orthogonal to each
other. 

We again write 
$$\ket{\phi} = \alpha\cdot \ket{\phi_{\parallel}} +
\beta\cdot\ket{\phi_{\perp}}$$
and we notice that $\lambda_0 = \alpha$, and 
$\ket{\psi_0} =\ket{\phi_\parallel}\otimes \ket{Z_T}^B$

Therefore the probability that Bob obtains all-zero in the measurement
is at least $|\lambda_0|^2 = |\alpha|^2\ge 1-\epsilon$.

After the measurement, and if is result is indeed all-zero, the state
will become
\begin{equation}
\ket{\psi} = {1\over\sqrt{|\lambda_0|^2 + |\lambda_1|^2}}
\cdot(\lambda_0\ket{\psi_0} + \lambda_1\ket{\psi_1})
\end{equation}
where $\ket{\psi_0}$ is in the diagonal subspace ${\cal H}^{\cal D}$ and
$\ket{\psi_1}$ is orthogonal to the ${\cal H}^{\cal D}$. 
The fidelity of $\ket{\psi}$ and ${\cal H}^{\cal D}$ is 
${{|\lambda_0|}\over{\sqrt{|\lambda_0|^2 + |\lambda_1|^2}}}$.

Now we can prove that the fidelity of $\ket{\psi}$ is at least
$1-\epsilon$:
\begin{eqnarray*}
\braket{\psi}{\Psi_N} & = & 
{\lambda_0\over\sqrt{|\lambda_0|^2 +
    |\lambda_1|^2}}\braket{\psi_0}{\Psi_N} +
{\lambda_1\over\sqrt{|\lambda_0|^2 +
    |\lambda_1|^2}}\braket{\psi_1}{\Psi_N}\\
& = & {{\braket{\lambda_0\psi_0}{\Psi_N}}\over\sqrt{|\lambda_0|^2 +
    |\lambda_1|^2}}\\
& = & {1\over\sqrt{|\lambda_0|^2 + |\lambda_1|^2}}\cdot
{{|\sum_{x\in X}\alpha_{x,x}|}\over{\sqrt{N}}} \\
& = & {{\braket{\phi}{\Psi_N}}\over\sqrt{|\lambda_0|^2 +
    |\lambda_1|^2}} \\
& \ge & {\braket{\phi}{\Psi_N}} \ge 1-\epsilon
\end{eqnarray*}
Essentially, the hash and compare protocol leaves the coefficients in
diagonal subspace untouched, and eliminates part of the
``off-diagonal'' coefficients. Therefore, after the re-normalization,
the coefficients in the diagonal subspace will not decrease, and thus
the fidelity of the output state is at least $1-\epsilon$.

\end{enumerate}

Now we estimate the magnitude of $\lambda_1$: we have
$$|\lambda_1|^2 = \sum_{x_A \ne x_B, x_A\oplus
  x_B\in Z}|\alpha_{x_A,x_B}|^2
$$
Notice that $\lambda_1$ is actually a random variable since the 
$r_0,r_1, ..., r_{t-1}$ are randomly chosen by Alice. Notice that each
pair $x_A\ne x_B$, we have
$$\prob_r[(x_A\oplus x_B)\bullet r = 0] = 1/2$$
and thus for random $r_0, r_1, ..., r_{s-1}$, the probability that all
$(x_A\oplus x_B)\bullet r_j$ results in 0 for $j=0, 1, ..., s-1$, is
$1/2^s$.

In other words, the expected value of $|\lambda_1|^2$ is
\begin{eqnarray*}
E[|\lambda_1|^2] & = &
E[\sum_{x_A \ne x_B, x_A\oplus x_B\in Z}|\alpha_{x_A,x_B}|^2] \\
& = & 
\sum_{x_A\ne x_B}\prob_{r_0, r_1, ..., r_{s-1}}[
x_A\oplus x_B \in Z]\cdot|\alpha_{x_A,x_B}|^2 \\
& = & {1\over{2^s}}\sum_{x_A\ne x_B}|\alpha_{x_A,x_B}|^2 \\
& \le & {\epsilon\over S}
\end{eqnarray*}
and thus by Markov Inequality, we have
$$\prob[|\lambda_1|^2\le {\epsilon\over\sqrt{S}}] \le
{1\over\sqrt{S}}$$
Therefore, with probability at least $1-{1\over\sqrt{S}}$, we have
$|\lambda_1|^2 \le {\epsilon\over\sqrt{S}}$. In that case, the
fidelity of $\ket{\psi}$ and the diagonal subspace is
\begin{eqnarray*}
F(\ket{\psi}, {\cal H}^{\cal D}) & = & 
{{|\lambda_0|^2}\over{{|\lambda_0|^2 + |\lambda_1|^2}}} \\
& \ge & 
{{{1-\epsilon}\over{1-\epsilon+ {\epsilon\over\sqrt{S}}}}} \\
&  = &
{1-{\epsilon\over{\sqrt{S}\cdot (1-\epsilon+
      {\epsilon\over\sqrt{S}})}}} \\
& \ge & {1-{{2\epsilon}\over{\sqrt{S}}}}
\end{eqnarray*} 
when $\epsilon < 1/2$.
\end{proof}

Now, we can put everything together: for a general state $\rho$, we
first apply the Hash and Compare protocol to 
$\rho$ to make it ``almost completely in'' the diagonal subspace
${\cal H}^{\cal D}$. Then we apply the Simple Scrambling protocol to 
enhance the fidelity. We describe the complete protocol in more
details: 

\begin{construction}[Complete Scrambling Protocol]
\label{construct:full-protocol}
The Complete Scrambling protocol is  parameterized by a quintuple:
$\langle N, K, W, L, S \rangle$, such that there exists
a Scrambling Permutation pair $\langle g_y(x), h_y(x)\rangle$ of
parameter $\langle N, K, W, L\rangle$, where $S$ is a power
of 2. 
The input to the protocol is a (mixed) state $\rho$ in space
${\cal H}_N^A\otimes {\cal H}_N^B$. The protocol also has an auxiliary
input $\Psi_T$, where $T = S \cdot K$. We can also write the
auxiliary input $\Psi_T$ as $\Psi_S \otimes \Psi_K$.
The steps are:

\begin{enumerate}
\item Alice and Bob engage in the Hash and Compare protocol, using the
  input state $\rho$ as the input, and part of the auxiliary
  input, $\Psi_S$ as the auxiliary input.
\item If the Hash and Compare protocol fails, Alice and Bob output
  $\fail$ and terminate.
\item If the Hash and Compare protocol succeeds, it will output a
  state $\sigma$. Alice and Bob then engage in the Simple Scrambling
  protocol, using $\sigma$ as the input and the other part of the
  auxiliary input, $\Psi_K$ as the auxiliary input.
\item If the Simple Scrambling protocol fails, Alice and Bob output
  $\fail$ and terminate.
\item If the Simple Scrambling protocol succeeds, a state $\tau$ will
  be output, and Alice and Bob output $\tau$.
\end{enumerate}
\end{construction}

It is obvious that the complete scrambling protocol can be realized
quantum-mechanically, and if the scrambling permutation used in the
protocol is an efficient one, and $L$ is a power of 2, the protocol
can be realized efficiently.

\begin{lemma}
\label{lemma:complete-scramble}
The Complete Scrambling protocol is a probabilistic conditional
successful GEPP with parameter 
\begin{math}
\langle N, SK, WK, \epsilon, ({4M\over N} +
{4\over\sqrt{S}})\epsilon, 2\epsilon + \sqrt{2\epsilon\over\sqrt{S}},
{1\over\sqrt{S}}\rangle 
\end{math}.
If the Simple Scrambling protocol used inside the complete protocol is
efficient, then so is the complete protocol.
\end{lemma}

\begin{proof}

To prove this lemma we need some claims about fidelity:
\begin{claim}[Monotonicity]
\label{claim:fidelity-monotone}
For any (mixed) states $\rho$ and $\sigma$ and any quantum operator
$\cal E$ (not necessarily unitary), we have
\begin{equation}
F({\cal E}(\rho), {\cal E}(\sigma)) \ge F(\rho, \sigma)
\end{equation}
\qed
\end{claim}
It is a well-known result~\cite{NC00}.

\begin{claim}[Triangle Inequality]
\label{claim:fidelity-triangle}
For any 3 pure states $\ket{A}$, $\ket{B}$ and $\ket{C}$ in the same
Hilbert space ${\cal H}$ such that
$F(\ket{A}, \ket{B}) = {1-\epsilon}$ and 
$F(\ket{A}, \ket{C}) = {1-\delta}$, where both $\epsilon$ and
$\delta$ 
are real numbers between $0$ and $1/2$.
then we have
$$F(\ket{B}, \ket{C}) \ge {1-2(\epsilon+\delta)}$$
\end{claim}
\begin{claim}[Relationship to Statistical Distance]
\label{claim:fidelity-prob}
Let $\rho$ and $\sigma$ be 2 mixed states in the Hilbert space
${\cal H}$ such that that $F(\rho, \sigma) = 1-\epsilon$.
Let $\cal E$ be an arbitrary quantum operation over
${\cal H}$ that ends with a measurement. We use $M_\rho$ and
$M_\sigma$ to denote the random variables describing the outcomes of
the measurement of ${\cal E}$ on input $\rho$ and $\sigma$,
respectively. Then the 
statistical distance between $M_\rho$ and $M_\sigma$ is at most 
$\sqrt{\epsilon}$.
\end{claim}
We prove Claim~\ref{claim:fidelity-triangle} and
Claim~\ref{claim:fidelity-prob}  in Appendix~\ref{app:fidelity}.

We first consider the case that the input state is a pure state
$\ket{\phi}$. By Lemma~\ref{lemma:hash-compare}, with probability at
least $1-\epsilon$, the Hash and Compare protocol will succeed. 
In the case it succeeds, the output state $\ket{\psi}$ will have a
fidelity at least $(1-{2\epsilon\over\sqrt{S}})$ with the
diagonal subspace  
${\cal  H}^{\cal D}$ with probability $1-{1\over\sqrt{S}}$. We define a
``good event'' to be the event that $\ket{\psi}$ has fidelity at least
$(1-{2\epsilon\over\sqrt{S}})$ with ${\cal H}^{\cal D}$. Then the
probability a good event happens is at least $1-{1\over\sqrt{S}}$. We
focus on the good events.
We write the normalized projection of $\ket{\psi}$ to 
${\cal H}^{\cal D}$ as $\ket{\psi_{\cal D}}$. So we have
$$F(\ket{\psi}, \ket{\psi_{\cal D}}) \ge {1-{2\epsilon\over\sqrt{S}}}$$

In other words, the fidelity of state $\ket{\psi}$ and the state 
$\ket{\psi_{\cal D}}$ is at least $1-{2\epsilon\over\sqrt{S}}$.
If Alice and Bob, instead of feeding $\ket{\psi}$, had fed
$\ket{\psi_{\cal D}}$ into the Simple Scrambling protocol, they would
have succeeded with probability at least $1-\epsilon$, and output a pure
state $\ket{\psi^E_{\cal D}}$ of fidelity at least 
$1-{2W\over  N}\epsilon$. However, since Alice and Bob don't feed
$\ket{\psi_{\cal D}}$ into the Simple Scrambling protocol, they don't
get $\ket{\psi^E_{\cal D}}$ back: rather they get a state
$\ket{\psi^E}$  if
they don't fail\footnote{It is easy to check that the Simple
  Scrambling protocol always outputs a pure state if the input state
  is pure.}. By the monotonicity of fidelity, we have that
\begin{equation}
\label{eqn:complete-scramble-1}
\braket{\psi^E}{\psi^E_{\cal D}} \ge \braket{\psi}{\psi_{\cal D}} \ge
1-{2\epsilon\over \sqrt{S}}
\end{equation}
Combining Equation~\ref{eqn:complete-scramble-1} with the fact that
\begin{math}
F(\ket{\psi^E_{\cal D}}, \ket{\Psi_{WK}}) \ge {1-{2W\over N}\epsilon}
\end{math}, 
we have, by Claim~\ref{claim:fidelity-triangle},
$$F(\ket{\psi^E}, \ket{\Psi_{WK}}) \ge {
1 - ({4W\over N} + {4\over\sqrt{S}})\epsilon}$$

We denote by $p$ the failing probability of the Simple Scrambling
protocol on input $\ket{\psi}$, and $p_{\cal D}$ the failing
probability on 
input $\ket{\psi_{\cal D}}$. Then we have, by
Claim~\ref{claim:fidelity-prob}, 
$$|p-p_{\cal D}| \le \sqrt{2\epsilon\over\sqrt{S}}$$

Putting things together, we have: with probability at least 
$1-2\epsilon - \sqrt{2\epsilon\over\sqrt{S}}$, the Complete Scrambling
protocol succeeds. In the case it succeeds, it outputs a
state $\ket{\psi^E}$ of fidelity at least 
$1-({4W\over N} + {4\over\sqrt{S}})\epsilon$ with probability at least
$1-{1\over\sqrt{S}}$.

Next we consider the case that the input state is a mixed state
$\rho$. We have $F(\rho) \ge 1-\epsilon$.
We write $\rho$  as an ensemble $\{p_i, \ket{\phi_i}\}$. For each
pure state $\ket{\phi_i}$, we assume that it has fidelity
$1-\epsilon_i$, and then by the linearity of fidelity, we have 
$\sum_{i}p_i\epsilon = \epsilon$. The analysis above works for each
pure state $\ket{\psi_i}$: for each pure state $\ket{\phi_i}$, with
probability  at least 
$1-2\epsilon_i - \sqrt{2\epsilon_i\over\sqrt{S}}$, the Complete
Scrambling protocol succeeds. In the case it succeeds, it 
outputs a state $\ket{\psi_i^E}$ of fidelity at least 
$1-({4W\over N} + {4\over\sqrt{S}})\epsilon_i$ with probability at
least $1-{1\over\sqrt{S}}$.
The fidelity $1-({4W\over N} + {4\over\sqrt{S}})\epsilon_i$
is a linear functions in $\epsilon_i$ , and 
$1-2\epsilon -  \sqrt{2\epsilon\over\sqrt{S}}$ is a convex function.
So overall, the Complete Scrambling protocol succeeds with probability
at least  
$1-2\epsilon - \sqrt{2\epsilon\over\sqrt{S}}$.
In the case it succeeds, it outputs a
state $\ket{\psi^E}$ of fidelity at least 
$1-({4W\over N} + {4\over\sqrt{S}})\epsilon$ with probability at least
$1-{1\over\sqrt{S}}$.
\end{proof}

Now we are ready to prove Theorem~\ref{thm:complete-scramble}.
\begin{proof}[Proof to Theorem~\ref{thm:complete-scramble}]
We simply combine Theorem~\ref{thm:construct-scramble} and
Lemma~\ref{lemma:complete-scramble} and choose $S=2^{2t}$.
\end{proof}


\section{Constructions  of Scrambling Permutations}
\label{sec:construct-scramble}
We discuss various constructions of Scrambling Permutations.

For a binary string $S = s_1s_2...s_n$, we define the \emph{left
  sub-string} and the \emph{right sub-string} of the string $S$ as
follows: 
\begin{eqnarray*}
\prefix(k, S) & = & s_1s_2....s_k \\
\postfix(k, S) & = & s_{n-k+1}s_{n-k+2}...s_n \\
\end{eqnarray*}
Obviously we have
$$\prefix(k,S) \circ \postfix(n-k,S) = S$$

The first construction is a very simple one, and it is very closely
related to a construction of universal hash functions.

\begin{construction}[Multiplication-table Scrambling Permutation]
\label{construct:multi-table}
We work in $GF_{2^n}$, where each element is a polynomial of degree at
most $n-1$, and can be written as
$$a_0 + a_1\cdot Z + \cdots + a_{n-1}\cdot Z^{n-1}$$
We identify each element with an $n$-bit
binary string in the most straight-forward way. We set $X = GF_{2^n}$
and $Y = GF_{2^n}^* = X \backslash \{\mathbf{0}\}$, where $\mathbf{0}$
is the additive identity in $GF_{2^n}$. We can pick an
arbitrary $l$, such that $1\le l < n$. Then we let 
$G = \{0,1\}^{l}$ and $H=\{0,1\}^{n-l}$. The functions are:
\begin{eqnarray*}
g_y(x) & = & \prefix(l, x \cdot y) \\
h_y(x) & = & \postfix(n-l, x \cdot y) \\
\end{eqnarray*}
and we have
$N=2^n$, $K= 2^n-1$, $L = 2^{l}$, and $M = 2^{n-l}$.
\end{construction}

Notice that a very common construction for universal hash functions
over $GF_{2^n}$ is $h_{y,z}(x) = x\cdot y + z$, and our construction
can be viewed as a sub-family of this universal hash family, by
setting $z=0$. Our construction here is \emph{not} a universal hash
function family, but is more efficient.

\begin{lemma}
The function pair given in Construction~\ref{construct:multi-table} is an
efficient Scrambling Permutation pair.
\end{lemma}
\begin{proof}
It is obvious that $\langle g_y(\cdot), h_y(\cdot)\rangle$ is a
permutation, since 
$$g_y(x)\circ h_y(x) = x \cdot y$$
is a permutation for $y\ne \mathbf{0}$.

Now let's prove that for any $x_1\ne x_2$, $\prob_{y}[h_y(x_1) =
h_y(x_2)]$ is always  the same. This is actually not hard: we have
$h_y(x_1) = h_y(x_2)$, iff 
$$(x_1 - x_2) \cdot y = 0 \;\mod \;(Z^{n-l})$$
There are exactly $2^l$ elements in $GF_N$ that are multiples of
$(Z^{n-l})$, and so there are exactly $2^l$ $y$'s that satisfy the
equation. However, one such $y$ is $\mathbf{0}$ and has to to be
excluded. So the probability is 
$p = (2^l-1)/(2^n-1)=(L-1)/(N-1)$. This is true
for every pair $x_1\ne x_2$.

Finally, both the permutation and its inverse can be
implemented efficiently (only field multiplication and inversion are
involved). So this scrambling permutation is efficient.
\end{proof}

A word about efficiency: it is desirable for us to construct families
of scrambling permutations of relatively small $K$ and $L$, as
compared $M$: In the Simple Scrambling protocol, where the
Scrambling Permutation is used, $N$
is the dimension of the input state that Alice and Bob try to
purify, which is normally fixed; $K$ is the dimension of maximally
entangled state Alice and 
Bob invests; $M$ is the ``yield'' of the protocol, or the
dimension of the output; $L$ is the dimension of the subspace Alice
and Bob discard. So the Simple Scrambling protocol invests about
$\log K$ perfect EPR pairs and discard about $\log L$ amount of
entanglement. For the Multiplication-table construction, $K$ is
almost as large as $N$, which is a disadvantage since 
Alice and Bob has to invest as many  perfect EPR pairs as the
imperfect ones they try to purify.
However, the $L$ in this construction is fully
adjustable, and it provides a nice trade-off between the yield Alice
and Bob wish to obtain and the fidelity of the output (the greater $L$
is, the less the yield is, and the higher fidelity the output has).

\remove{
Apparently we hope to invest as few fresh qubits as possible. In other
words, we hope to minimize $M$ while still maintaining the property of
the protocol. Fortunately the most 2 important parameters, the success
probability and the fidelity, are both independent to $M$. Thus if we
can find a good scrambling permutation, we are all right. A good
scrambling permutation should have the maximum ``scrambling''
possible: the function pair $\langle g_y(x), h_y(x)\rangle$ should be
such that the output size of $g_y(x)$ is very small, and most
information is concentrated in $h_y(x)$.

Here is a trivial way to ``extend'' a scrambling permutation that
doesn't work: take any existing $\langle g_y(x), h_y(x)\rangle$,
define a new permutation pair as follows: take any $x\circ x'$ and
$y$', we define $g_y'(x\circ x') = g_y(x)\circ x'$ and $h_y'(x\circ
x') = h_y(x)$. In this way we can arbitrarily extend the size of $X$,
and therefore effectively decreasing the relative size of
$Y$. However, this scheme doesn't buy us anything since all the
excessive information goes to $G$, which will be measured and are
useless.
}

Below is another construction:

\begin{construction}[Linear Function Scrambling Permutation]
\label{construct:linear-func}
We work in $GF_{2^n}$, and let $X=GF_{2^n}\times GF_{2^n}$. Therefore
each element in $X$ is represented by $\langle x_0, x_1\rangle$. We
let $Y= GF_{2^n} \cup \{\perp\}$, where $\perp$ is a special symbol.

Both functions $g_y(\langle x_0, x_1\rangle)$ and 
$h_y(\langle x_0, x_1\rangle)$ output elements in $GF_{2^n}$
and the actual functions are defined as follows:

\begin{eqnarray*}
g_y(\langle x_0, x_1\rangle) & = & 
\left \{
\begin{array}{lll}
x_0 & & \mbox{\sf if $y\in GF_{2^n}$} \\
& & \\
x_1 & & \mbox{\sf if $y = \perp$} \\
\end{array}
\right. \\
h_y(\langle x_0, x_1\rangle) & = & 
\left \{
\begin{array}{lll}
x_0 \cdot y + x_1  & & \mbox{\sf if $y\in GF_{2^n}$} \\
& & \\
x_0 & & \mbox{\sf if $y = \perp$} \\
\end{array}
\right. \\
\end{eqnarray*}
and we have
$N = 2^{2n}$, $K= 2^n+1$, $M = 2^n$, and $L= 2^n$.
\end{construction}

\begin{lemma}
The function pair given in Construction~\ref{construct:linear-func} is
an efficient Scrambling Permutation pair.
\end{lemma}
\begin{proof}
It is easy to verify that for any $y$, $g_y(\langle x_0, x_1\rangle)
\circ h_y(\langle x_0, x_1\rangle)$ is a permutation.

Next we prove the scrambling property: for any pair of inputs
$x = \langle x_0, x_1 \rangle$ and 
$x' = \langle x_0', x_1' \rangle$:
\begin{itemize}
\item
If $x_0 \ne x_0'$, then the unique 
$y = (x_1 - x_1') \cdot (x_0 - x_0')^{-1}$ makes 
$ h_y(\langle x_0, x_1\rangle) =  h_y(\langle x_0', x_1'\rangle)$.
\item
If $x_0 = x_0'$, then the unique $y = \perp$ makes 
$ h_y(\langle x_0, x_1\rangle) =  h_y(\langle x_0', x_1'\rangle)$.
\end{itemize}
Finally, it is easy  both the permutation and its can be computed
efficiently, and thus the linear function construction is an efficient
Scrambling Permutation pair.
\end{proof}

In this construction, $K$ is about the square root of $N$, which is
much better than the Multiplication-table construction. However, $L$
is fixed, and we don't have the flexibility as in the
Multiplication-table construction. However, we can extend this
construction to a class of Scrambling Permutations, and resolve the
flexibility problem.

\begin{construction}[Extended Linear Function Scrambling Permutation]
\label{construct:linear-func-ext}
We work in $GF_{2^n}$, and let
$X=GF_{2^n}^d$, where $d$ is an integer. Therefore each
element in $X$ is represented by a $d$-tuple $\langle x_0, x_1, ..., 
x_{d-1}\rangle$. We let $Y= \bigcup_{k=0}^{d-1} GF_{2^n}^d$, where we
define $GF_{2^n}^0 = \{\perp\}$.

The function $g_y(\langle x_0, x_1, ... , x_{d-1}\rangle)$ outputs an
element in $GF_{2^n}$ and the function  
$h_y(\langle x_0, x_1, ..., x_{d-1} \rangle)$ output a $(d-1)$-tuple
in $GF_{2^n}$:
For any $y\in Y$, we write $y = \langle y_0, y_1, ...,
y_{k-1}\rangle$, where $0\le k < d$.
\begin{eqnarray*}
g_y(\langle x_0, x_1, ..., x_{d-1}\rangle) & = & 
x_{k} \\
h_y(\langle x_0, x_1, ..., x_{d-1}\rangle) & = & 
\langle 
x_0 + x_k\cdot y_0, 
x_1 + x_k\cdot y_1, ... , 
x_{k-1} + x_k\cdot y_{k-1}, 
x_{k+1}, x_{k+2}, ... , x_{d-1}
\rangle 
\end{eqnarray*}
and we have
$N = 2^{dn}$, $K= {{2^{dn}-1}\over{2^n-1}}$, $M = 2^{(d-1)n}$, and 
$L = 2^n$. 
\end{construction}
Here is a concrete example for $d=4$:
\begin{center}
\begin{tabular}{|l||c|c@{$\;,\;$}c@{$\;,\;$}c|}
\hline
$y$ & $g_y(x)$ & \multicolumn{3}{c|}{$h_y(x)$}\\
\hline
$y = \perp$ & $x_0$ & $x_1$ & $x_2$ & $x_3$ \\
\hline
$y= \langle y_0 \rangle$ & $x_1$ &  $x_0 +x_1\cdot y_0$ & $x_2$ 
& $x_3$ \\
\hline
$y=\langle y_0, y_1 \rangle$ & $x_2$ & 
$x_0+x_2\cdot y_0$ & $x_1 + x_2\cdot y_1$ & $x_3$ \\
\hline
$y=\langle y_0, y_1, y_2 \rangle$ & $x_3$ & 
$x_0+x_3\cdot y_0$ & $x_1 + x_3\cdot y_1$ & $x_2 + x_3\cdot y_2$ \\
\hline
\end{tabular}
\end{center}

\begin{lemma}
The function pair defined in
Construction~\ref{construct:linear-func-ext} is an efficient
Scrambling Permutation pair.
\end{lemma}
\begin{proof}
The permutation property is obvious, and it is easy to see that both
the permutation and its inverse can be computed efficiently.

Now the scrambling property: given any pair $x = 
\langle x_0, x_1, ..., x_{d-1}\rangle$ and
$x' = \langle x_0', x_1', ..., x_{d-1}'\rangle$. we show that there is
always a unique $y$ such that $h_y(x) = h_y(x')$. We define $k$ to be
the largest index such that $x_k\ne x_k'$. Then for $y\in GF_{2^n}^l$,

\begin{enumerate}
\item
If $l< k$,
then the $k$-th entry in $h_y(x)$ is $x_k$, and it is different from
the $k$-th entry in $h_y(x')$, which is $x_k'$; 
\item 
If $l = k$, we are effectively solving a linear system:
\begin{eqnarray*}
x_0 + x_k \cdot y_0 & = & x_0' + x_k' \cdot y_0 \\
x_1 + x_k \cdot y_1 & = & x_1' + x_k' \cdot y_1 \\
& ... & \\
x_{k-1} + x_k \cdot y_{k-1} & = & x_{k-1}' + x_k' \cdot y_{k-1} \\
\end{eqnarray*}
and it has a unique solution
\begin{eqnarray*}
y_0 & = & (x_0 - x_0') \cdot (x_k - x_k')^{-1} \\
y_1 & = & (x_1 - x_1') \cdot (x_k - x_k')^{-1} \\
& ... &  \\
y_{k-1} & = & (x_{k-1} - x_{k-1}') \cdot (x_k - x_k')^{-1} \\
\end{eqnarray*}

\item
If $l>k$, the $k-$th entry of $x$ is $x_k + y_k\cdot x_{k+1}$, and it
is different from  
the $k$-th entry of $x'$, which is $x_k' + y_k\cdot x_{k+1}$, since
$x_k \ne x_k'$, while $x_{k+1} = x_{k+1}'$.
\end{enumerate}

So there exists a unique $y\in Y$ such that $h_y(x) = h_y(x')$.
\end{proof}

The extended linear function construction gives a class of Scrambling
Permutations of different parameters: for a fixed $N$, we can pick a
construction such that $K$ is about $N^{(d-1)/d}$ and $L$ is about
$N^{1/d}$ for any integer $d$. When $d=2$, the extended linear
function construction becomes the linear function construction. So we
get back some flexibility: not only in $K$, but also in $L$. 

Of course, one question is: how good are our constructions in terms of
the size of $K$ and $L$ as compared to $N$? We hope $K$ and $L$ are as
small as possible, and how small can they be? We have the following
theorem which essentially says that the Extended Linear Function
construction is optimal in terms of the size of $K$ and $L$.

\remove{
The extended linear function example gives a family of constructions
of the scrambling permutations. For them, we have $|X| = 2^{dn}$,
$|Y| = \sum_{k=0}^{d-1}2^{kn} = (2^{dn}-1)/(2^n-1)$, and $|G| =
2^n$. Remember we want to make $|Y|$ and $|G|$ as small as possible as
compared to $|X|$, and in this sense the extended linear function
construction is better.
}

\begin{theorem}
\label{thm:scramble-perm-bound}
Let $\langle g_y(x), h_y(x)\rangle$ be a
scrambling permutation pairs of parameter $\langle N, K, M, L\rangle$.
We have $N \le KL$.
\end{theorem}
\begin{proof}
First, by Theorem~\ref{thm:scramble-perm-collision}, we know that
the collision probability $p = (L-1)/(N-1)$.

Recall that $p$ is the probability that a random $y\in Y$ satisfies
$h_y(x_1) = h_y(x_2)$, and thus it is at least $1/K$.
Therefore we have 
$${1\over{K}}\le {{L-1}\over{N}-1}$$ 
or 
$$K \ge {{N-1}\over{L}-1} \ge {{N}\over{L}}$$ 
\end{proof}

It is easy to see that the Extended Linear Function construction
achieves this bound asymptotically.

We summarize the 3 constructions in the following table, which
essentially proves Theorem~\ref{thm:construct-scramble}.

\begin{center}
\begin{tabular}{|l||c|c|c|c|l|l|}
\hline
\bf Construction & $N$ & $K$ & $M$ & $L$ & Comments \\
\hline
Multiplication-table &
$2^n$ & $2^n-1$ & $2^{n-l}$ & $2^{l}$ & 
Fully adjustable $L$, not optimal\\

Linear Function & 
$2^{2n}$ & $2^n+1$ & $2^n$ & $2^n$ & Minimal $K$ among all
constructions, optimal, inflexible $L$ \\

Extended Linear Function &
$2^{dn}$ & ${{2^{dn}-1}\over{2^n-1}}$ & $2^{(d-1)n}$& $2^n$
& Optimal, flexible $K$ and $L$ \\
\hline
\end{tabular}
\end{center}

\section{Proofs to Two Claims About Fidelity}
\label{app:fidelity}

We give the proofs to 2 claims about fidelity that are used in this paper.

\begin{proof}[Proof to Claim~\ref{claim:fidelity-triangle}]
Notice that $\ket{A}$, $\ket{B}$ and $\ket{C}$ are vectors in 
$\cal H$. We denote the angle between $\ket{A}$ and $\ket{B}$ by
$\theta_{AB}$, and define $\theta_{BC}$, and $\theta_{CA}$
accordingly. Then it is easy to see (by the triangle inequality),
that $\theta_{BC} \le \theta_{AB} + \theta_{AC}$. 
It is also easy to see that
$\cos\theta_{AB} = \braket{A}{B} = \sqrt{1-\epsilon}$
and
$\cos\theta_{AC} = \braket{A}{C} = \sqrt{1-\delta}$
Therefore, we have
\begin{eqnarray*}
\braket{B}{C} & = & \cos\theta_{BC} \\
& \ge & \cos(\theta_{AB} + \theta_{AC}) \\
& = & \cos\theta_{AB}\cos\theta_{AC} - \sin\theta_{AB}\sin\theta_{AC}
\\
& = & \sqrt{(1-\epsilon)(1-\delta)} - \sqrt{\epsilon\delta} \\
& \ge & \sqrt{1-2(\epsilon+\delta)} 
\end{eqnarray*}
where the last step is a simple algebraic deduction.

\end{proof}

\begin{proof}[Proof to Claim~\ref{claim:fidelity-prob}]
We use $D(\rho, \sigma)$ to denote the trace distance between $\rho$
and $\sigma$, and we have\cite{NC00}
$$D(\rho, \sigma) \le \sqrt{1-F(\rho, \sigma)} = \sqrt{\epsilon}$$
However the statistical distance between $M_\rho$ and $M_\sigma$ is
bounded by $D(\rho, \sigma)$, which is bounded by $\sqrt{\epsilon}$.
\end{proof}

\end{document}